\documentstyle[11pt,aaspp4]{article}

\def\la{\mathrel{\hbox{\rlap{\hbox{\lower4pt\hbox{$\sim$}}}\hbox{$<$}}}}
\def\ga{\mathrel{\hbox{\rlap{\hbox{\lower4pt\hbox{$\sim$}}}\hbox{$>$}}}}

\slugcomment{Accepted to {\it The Astrophysical Journal}}

\begin{document}

\title{Light Curves and Spectra of Dust Echoes From Gamma-Ray Bursts and 
their Afterglows:  Continued Evidence that GRB 970228 is Associated with a 
Supernova}

\author{Daniel E. Reichart\altaffilmark{1,2}}

\altaffiltext{1}{Department of Astronomy, California Institute of Technology, 
Mail Code 105-24, 1201 East California Boulevard, Pasadena, CA 91125}
\altaffiltext{2}{Hubble Fellow}

\begin{abstract}

The relative brightening and reddening of the optical afterglows of GRB 
970228 and GRB 980326 around 20 -- 30 days after these bursts have been 
attributed to supernovae, which are red (blueward of their spectral peak), 
and which peak in time after $\approx 20(1+z)$ days.  However, this direct 
evidence for a GRB/SN connection has recently been challenged.  It has been 
suggested that the late afterglows of these bursts can be explained by dust 
echoes, of which we consider two cases:  (1) the scattering of light from the 
afterglow (the forward shock), the optical flash (the reverse shock), and/or 
optical light from the burst itself by a shell of dust at a radius $R$ from 
the progenitor, and (2) the absorption and thermal re-emission of this light 
by this shell of dust.  In this paper, we model and compute dust echo light 
curves and spectra for both cases.  Although the late-time afterglow of GRB 
980326 was not sufficiently well sampled to rule out a dust echo description, 
we find that the late-time afterglow of GRB 970228 cannot be explained by a 
dust echo.  

\end{abstract}


\section{Introduction}

Bloom et al. (1999) first identified a second component to the optical 
afterglow of GRB 980326.  This second component grew brighter than the 
extrapolation of the early afterglow (which faded as a power law) around 1 
week after the burst, and peaked around 20 -- 30 days after the burst before 
fading away.  Rough spectra taken 2.4 and 28 days after the burst show that 
the second component was significantly redder than the early afterglow.  
Bloom et al. (1999) attributed this behavior to a supernova at a redshift of 
$z \sim 1$.  Supernovae peak after about $20(1+z)$ days and are significantly 
redder than afterglows (blueward of the supernova's spectral peak).  
Unfortunately, neither a redshift nor multiband photometry is available for 
this afterglow.

Reichart (1999) identified a similar second component to the optical 
afterglow of GRB 970228, the first burst for which an afterglow was detected. 
 Reichart (1999) and Galama et al. (2000) showed that the light curve of this 
second component is consistent with the expectation of a supernova at the 
spectroscopically-measured redshift of this burst (z = 0.695; Bloom, Djorgovski, 
\& Kulkarni 2001).  More convincingly, VRIJK photometry of the second component 
taken 
near its peak showed that its spectral flux distribution is also consistent 
with the expectation of a supernova at this redshift:  (1) the spectral flux 
distribution is very red blueward of the spectral peak and very blue redward 
of the spectral peak, and (2) the spectral peak is at the frequency of the 
spectral peak of a supernova at $z \approx 0.7$ (Figure 1; Reichart 1999; 
Galama et al. 2000; Reichart, Castander, \& Lamb 2000; Reichart, Lamb, \& 
Castander 2000).  This is the best evidence to date for the GRB/SN connection.

The GRB/SN connection has recently been challenged by alternative 
explanations of these observations, called dust echoes (Waxman \& Draine 
2000; Esin \& Blandford 2000; see also Lamb 1997, 2000).  If the progenitors 
of bursts are massive stars (see Lamb \& Reichart 2000 for a review of the 
indirect evidence that bursts are associated with star-forming regions), due 
to the progenitor's wind, the density of the surrounding medium should fall 
off as $n(r) \propto r^{-2}$ out to the radius $R_w$ of the wind termination 
shock (Chevalier \& Li 1999).  At $R_w$, the density should jump by a factor 
of $\approx 4$, and beyond $R_w$, the density should remain approximately 
constant out to great distances (Chevalier \& Li 1999).  Dust in the vicinity 
of the burst should be sublimated by UV light from the optical flash (the 
reverse shock; Waxman \& Draine 2000).  Let $R$ be the radius out to which 
dust is sublimated.  If $R < R_w$, and if the dust is sublimated isotropically, 
the dust should occupy a relatively thin shell at $\approx R$ due to the 
rapidly decreasing density of the circumburst medium with distance.  Light 
that is emitted by the afterglow (the forward shock) from a distance that is 
much less than $R$ at an angle $\theta$ relative to the line of sight, is 
scattered (Esin \& Blandford 2000) or absorbed and thermally re-emitted 
(Waxman \& Draine 2000; Esin \& Blandford 2000) by dust at $r \approx R$ into 
the line of sight, and arrives at the observer, arrives with an 
observer-frame time delay of 
\begin{equation}
t = \frac{r(1+z)}{c}(1- \cos{\theta})
\label{delay}
\end{equation}
relative to light that was emitted into the line of sight ($\theta = 0$) and 
not scattered or absorbed.  This dust and light geometry is depicted in the 
top panel of Figure 2.  Waxman \& Draine (2000) and Esin \& Blandford (2000) 
find that this results in secondary components to the afterglow, called dust 
echoes, that may peak about 20 -- 30 days after the burst, and that may be 
significantly redder than the early afterglow, i.e., that the late-time 
afterglows of GRB 970228 and GRB 980326 may in fact be due to dust echoes, 
and not supernovae.

There are a number of problems with this dust and light geometry.  First of 
all, $R \approx 12$ pc (Waxman \& Draine 2000); however, $R_w$ is probably 
only a few parsecs (Chavalier \& Li 2000).  
Consequently, the dust would probably not occupy a thin shell. 

Secondly and more importantly, UV light from the optical flash will only 
sublimate dust within $\theta_{jet}$ of the line of sight, where $\theta_{jet}$ 
is the initial half-opening angle of the jet. (The Lorentz factor $\gamma$ of 
the shocked material in the jet is much greater than $\theta_{jet}^{-1}$ when 
the optical flash occurs.)  If the shock is adiabatic, 
\begin{equation}
\gamma \approx 
6.1(1+z)^{3/8}\left(\frac{n}{\rm{1\,cm}^3}\right)^{-1/8} 
\left(\frac{E}{10^{52}\rm{\,erg}}\right)^{1/8} 
\left(\frac{t}{\rm{1\,dy}}\right)^{-3/8},
\end{equation}
where $n$ is the number density of gas atoms in the circumburst medium, $E$ 
is the isotropic-equivalent energy of the shock, and $t$ is observer-frame 
time (e.g., Sari, Piran, \& Narayan 1998).  Let $n = 
n(\rm{1\,pc})(r/\rm{1\,pc})^{-2}$.  Since $r \approx 4\gamma^2ct/(1+z)$ 
(e.g., Sari, Piran, \& Narayan 1998), 
\begin{equation}
\gamma \approx 
2.2(1+z)^{1/4}\left[\frac{n(\rm{1\,pc})}{\rm{1\,cm}^3}\right]^{-1/4} 
\left(\frac{E}{10^{52}\rm{\,erg}}\right)^{1/4} 
\left(\frac{t}{\rm{1\,dy}}\right)^{-1/4}.
\label{gamma}
\end{equation}
The afterglow emits light into a half-opening angle $\theta_{col} \approx 
(\theta_{jet}^2 + \gamma^{-2})^{1/2}$.  Given the rate at which $\gamma$ 
decreases with time, $\theta_{col}$ quickly grows significantly greater than 
$\theta_{jet}$:  For example, $\theta_{col} > 1.1\theta_{jet}$ after only a few 
minutes if 
$\theta_{jet} \approx 0.1$:\footnote{When $\gamma \approx \theta_{jet}^{-1}$, 
the jet begins to expand laterally, and the afterglow fades much more rapidly 
(e.g., Panaitescu, M\'esz\'aros, \& Rees 1998).  Applying Equation 
(\ref{gamma}) to the afterglows of GRB 990510 (e.g., Stanek et al. 1999; 
Harrison et al. 1999), GRB 991216 (Halpern et al. 2000), GRB 000301c 
(e.g., Rhoads \& Fruchter 2000; Berger et al. 2000), and GRB 000926 (Price et 
al. 2001), we find that 
$\theta_{jet} \approx 0.17[n(\rm{1\,pc})/\rm{1\,cm}^3]^{1/4}$, 
$0.14[n(\rm{1\,pc})/\rm{1\,cm}^3]^{1/4}$, 
$0.38[n(\rm{1\,pc})/\rm{1\,cm}^3]^{1/4}$, and 
$0.18[n(\rm{1\,pc})/\rm{1\,cm}^3]^{1/4}$, respectively.}
\begin{equation}
t \ga 
14(1+z)\left[\frac{n(\rm{1\,pc})}{\rm{1\,cm}^3}\right]^{-1} 
\left(\frac{E\theta_{jet}^2}{10^{52}\rm{\,erg}}\right) 
\left(\frac{\theta_{jet}}{0.1}\right)^2{\rm\,min}.
\end{equation}
Light emitted at $\theta > \theta_{jet}$ would be scattered, absorbed, and 
thermally re-emitted by dust at radii that are substantially smaller than 
$R$.  Due to the substantially higher density of dust at these radii, this 
light would dominate the dust echo.  Radiative shocks yield similar results.  
This more realistic dust and light geometry is depicted in the bottom panel of 
Figure 2.  

In this paper we consider a dust and light geometry of which the dust and light 
geometry depicted in the bottom panel of Figure 2 is a limiting case.  
High-energy light from the progenitor may destroy dust (e.g., via electrostatic 
disruption; Draine \& Salpeter 1979; Waxman \& Draine 2000) out to a 
smaller radius.  Similarly, irregularities in the progenitor's wind may destroy 
dust in shocks.  In either case, the dust would be destroyed isotropically.  Let 
$R$ instead be the radius out to which dust is destroyed by the progenitor, 
and let $R_{jet}$ be the radius out to which dust is destroyed by the burst, 
the optical flash, and/or the afterglow, by whatever dust destruction 
mechanisms.  If $R_{jet}$ is even twice as great as $R$, and if the jet is 
oriented along the line of sight, the dust echoes should brighten 
substantially at a time given by Equation (\ref{delay}) with $r \approx R$ 
and $\theta \approx \theta_{jet}$, and dim substantially at a time given by 
Equation (\ref{delay}) with $r \approx R$ and $\theta \approx 
\pi-\theta_{jet}$, since we assume that the jet is bipolar.  If $\theta_{jet} 
\ll 1$, then the dust echoes ``turn on'' at
\begin{equation}
t_{on} \approx 
6.0(1+z)\left(\frac{R}{\rm{1\,pc}}\right)\left(\frac{\theta_{jet}}{0.1}\right)^2
\rm{\,dy},
\end{equation} 
and ``turn off'' at 
\begin{equation}
t_{off} \approx 6.5(1+z)\left(\frac{R}{\rm{1\,pc}}\right)\rm{\,yr\,} - t_{on},
\end{equation} 
though they may fade earlier than $t_{off}$ due to other effects (see \S 2 
and \S 3).  The jet may be oriented as much as $\approx \theta_{jet}$ from 
the line of sight (any more and we would not have seen the burst).  
Consequently, the dust echoes may ``turn on'' half of the way as early as $t 
\approx 0$ and all of the way as late as $t \approx 4t_{on}$.  The ``turn 
off'' time would be similarly affected.  The dust and light geometry depicted in 
the bottom panel of Figure 2 is recovered in the limit that $R \rightarrow 0$.  

Another dust and light geometry that we consider in this paper is that light
from the optical flash, as well as optical light that may be associated with the
burst itself, may result in dust echoes.  Consequently, dust echoes may be a way 
to measure the brightness of the optical flash and/or the optical
brightness of the burst.

In this paper, we model and compute dust echo light curves and spectra for 
both $R_{jet} >$ a few times $R$, and $R_{jet} < R$ (in which case dust is 
destroyed only isotropically).  Any other case would be fine tuning.  We also 
model and compute dust echoes resulting from a pulse of light (i.e., the 
optical flash or optical light associated with the burst), and an afterglow 
light curve.  Finally, we model and compute dust echo light curves and 
spectra for both scattered light (\S 2), and absorbed and thermally 
re-emitted light (\S 3).  We find that both types of dust echoes have light 
curves and spectra that can be distinguished from each other, and from the 
light curves and spectra of supernovae.  In \S 4, we re-examine the late-time 
afterglow of GRB 970228, and find that it cannot be explained by a dust echo. 
 We draw conclusions in \S 5.

\section{Dust Echoes from Scattered Light}

\subsection{Dust Echoes from a Scattered Isotropic Pulse of Light}

Before modeling the general case of the dust echo of collimated light that is 
emitted as a function of time (e.g., an afterglow), we first model the 
simpler case of the dust echo of an isotropic pulse of light (e.g., an 
optical flash or optical light associated with the burst) that is emitted 
from a distance that is much less than $R$.  Consider the light that is 
emitted at an angle $\theta$ relative to the line of sight, is scattered by 
dust at $r$ into the line of sight, and arrives at the observer.  The optical 
depth at observer-frame frequency $\nu$ along this path is given by:
\begin{equation}
\tau_{\nu (1+z)}(r,\theta) = \int_0^r n(r,\theta)\sigma_{\nu (1+z)}^{tot}dr + 
\int_{r\cos{\theta}}^{\infty}n(r,\theta)\sigma_{\nu 
(1+z)}^{tot}d(r\cos{\theta}),
\end{equation}
where $n(r,\theta)$ is the number density of dust particles at $r$ and 
$\theta$, and $\sigma_{\nu (1+z)}^{tot} = \sigma_{\nu (1+z)}^{abs} + 
\sigma_{\nu (1+z)}^{sc}$ is the total (absorption plus scattering) cross 
section of these particles.  The number density of dust particles at $r < 
R_w$ is given by (\S 1):
\begin{equation}
n(r,\theta) = \cases{0 & $(r < R)$ \cr
0 & $(r < R_{jet}$ and $0 < \theta < \theta_{jet}$ or $\pi-\theta_{jet} < 
\theta < \pi)$ \cr 
n_R(r/R)^{-2} & $(R < r < R_w$ and $f\theta_{jet} < \theta < 
\pi-f\theta_{jet})$ \cr
n_R(r/R)^{-2} & $(R_{jet} < r < R_w$ if $R < R_{jet} < R_{w})$},
\label{density}
\end{equation}
where (\S 1)
\begin{equation}
f = \cases{0 & $(R_{jet} < R)$ \cr 
1 & $(R_{jet} >$ a few times $R)$}.
\end{equation}
Assuming that $R_w >$ a few times $R$, we find that 
\begin{equation}
\tau_{\nu (1+z)}(r,\theta) = \cases{\tau_{\nu 
(1+z)}\left[1+\frac{R}{r}\left(\frac{\theta-f\theta_{jet}}{\sin{\theta}} - 
1\right)\right] & $(f\theta_{jet} < \theta < \frac{\pi}{2}$ or $r\sin\theta > 
R)$ \cr 
\tau_{\nu 
(1+z)}\left[1+\frac{R}{r}
\left(\frac{\theta-f\theta_{jet}-2\cos^{-1}\frac{r\sin\theta}{R}}{\sin{\theta}} 
- 1\right)\right] & $(\frac{\pi}{2} < \theta < 
\pi-f\theta_{jet}$ and $r\sin\theta < R)$ \cr
\tau_{\nu (1+z)}(3-\frac{2R}{r}) & $(\theta = \pi$ and $f = 0)$},
\label{tau}
\end{equation}
where $\tau_{\nu (1+z)} = n_R\sigma_{\nu}^{tot}R$ is the optical depth along 
any line of sight if $f = 0$, or any line of sight between $\theta_{jet} < 
\theta < \pi-\theta_{jet}$ if $f = 1$.  

The probability that light takes this path, i.e., that light that is emitted 
within $d\Omega$ of $\theta$ makes it to within $dr$ of $r$ without being 
scattered or absorbed, is scattered within $dr$ of $r$ into the line of 
sight, and makes it to the observer without being scattered or absorbed, is 
given by:
\begin{equation}
dP_{\nu (1+z)}^{sc}(r,\theta) = e^{-\tau_{\nu 
(1+z)}(r,\theta)}n(r,\theta)\frac{d\sigma_{\nu (1+z)}^{sc}}{d\Omega}d\Omega dr.
\end{equation}
We adopt the same differential dust scattering cross section as do Esin \& 
Blandford (2000; White 1979):
\begin{equation}
\frac{d\sigma_{\nu (1+z)}^{sc}}{d\Omega} = \frac{\epsilon_{\nu 
(1+z)}\sigma_{\nu (1+z)}^{tot}}{4\pi} \frac{1-\bar{\mu}_{\nu 
(1+z)}^2}{[1+\bar{\mu}_{\nu (1+z)}^2-2 \bar{\mu}_{\nu (1+z)}\mu]^{3/2}},
\label{scat}
\end{equation}
where $\mu = \cos{\theta}$, $\bar{\mu}_{\nu (1+z)}$ is the mean value of 
$\mu$ at $\nu(1+z)$ for the dust particles, and $\epsilon_{\nu (1+z)} = 
\sigma_{\nu (1+z)}^{sc}/\sigma_{\nu (1+z)}^{tot}$ is the mean albedo at 
$\nu(1+z)$ for these particles.  We adopt $\bar{\mu}_{\nu (1+z)} = 
\epsilon_{\nu (1+z)} = 0.5$ (White 1979).  The solid angle across which this 
light is scattered is $d\Omega = 
2\pi\sin{\theta}d\theta$.  Consequently,
\begin{equation}
dP_{\nu (1+z)}^{sc}(r,\theta) = \frac{3\tau_{\nu 
(1+z)}}{2}\frac{n(r,\theta)}{n_R} \frac{e^{-\tau_{\nu 
(1+z)}(r,\theta)}\sin{\theta}d\theta}{(5-4\cos{\theta})^{3/2}} 
d\left(\frac{r}{R}\right).
\label{prob}
\end{equation}
Using Equation (\ref{delay}), we rewrite Equation (\ref{prob}) as a function 
of $r$ and $t$:
\begin{equation}
dP_{\nu (1+z)}^{sc}(r,t) = \frac{3c\tau_{\nu 
(1+z)}}{2R(1+z)}\frac{n(r,\theta)}{n_R}\left(\frac{r}{R}\right)^{-1}
\frac{e^{-\tau_{\nu (1+z)}(r,\theta)}}{(5-4\cos{\theta})^{3/2}} 
d\left(\frac{r}{R}\right)dt,
\end{equation}
where $\theta$ is a function of $r$ and $t$ given by Equation (\ref{delay}).

Let $F_{\nu}dt$ be the fluence of the pulse of light at $\nu$.  Ignoring 
light that multiply scatters into the line of sight, the light curve of the 
dust echo is then given by:
\begin{eqnarray}
F^E_{\nu}(t) & = & \int\frac{F_{\nu}}{e^{-(1-f)\tau_{\nu (1+z)}}}dP_{\nu 
(1+z)}^{sc}(r,t) \\
& = & \frac{3c\tau_{\nu 
(1+z)}}{2R(1+z)}\int_r^{\infty}P(r')\left(\frac{r'}{R}\right)^{-3} 
\frac{e^{-[\tau_{\nu (1+z)}(r',\theta)-(1-f)\tau_{\nu 
(1+z)}]}}{(5-4\cos{\theta})^{3/2}}d\left(\frac{r
'}{R}\right) F_\nu dt,
\label{pulse}
\end{eqnarray}
where 
\begin{eqnarray}
P(r') = \cases{0 & $[r' < r(\pi-f\theta_{jet})]$ \cr
1 & $[r(\pi-f\theta_{jet}) < r' < r(f\theta_{jet})]$ \cr
0 & $[r' > r(f\theta_{jet})]$ },
\label{p}
\end{eqnarray}
\begin{equation}
r = \cases{R & $[t < t(\pi-f\theta_{jet})]$ \cr 
r(\pi-f\theta_{jet}) & $[t > t(\pi-f\theta_{jet})]$},
\end{equation}
and
\begin{equation}
\theta = \cases{\theta(R) & $[t < t(\pi-f\theta_{jet})]$ \cr 
\pi-f\theta_{jet} & $[t > t(\pi-f\theta_{jet})]$},
\end{equation}
and where
\begin{equation}
r(\theta) = \frac{ct}{(1-\cos{\theta})(1+z)},
\end{equation}
\begin{equation}
\theta(r) = \cos^{-1}\left[1-\frac{ct}{r(1+z)}\right],
\end{equation}
and
\begin{equation}
t(\theta) = \frac{(1-\cos{\theta})R(1+z)}{c}.
\end{equation}
If $f = 0$, or if $f=1$ and $\theta_{jet} \ll {\rm min}\{\theta, 
{(\theta/\tau)(r/R)}\}$ [Equations (\ref{p}) and (\ref{tau}), respectively], 
then Equation (\ref{pulse}) is approximately given by:
\begin{equation}
F^E_{\nu}(t) \approx \cases{A\left(\frac{r}{R}\right)^{-2} & $(t < t_{sc}$ 
and $t < t_{\tau})$ \cr
A\left(\frac{r}{R}\right)^{-2}\left(\frac{t}{t_{sc}}\right)^{-3/2}\left[4-3\left
(\frac{t}{t_{sc}}\right)^{-1/2}\right] & $(t_{sc} < t < t_{\tau})$ \cr
A\left(\frac{r}{R}\right)^{-2} \left(\frac{t_{\tau}}{t_{sc}}\right)^{-3/2} 
\left(\frac{t}{t_{\tau}}\right)^{-7/4} 
\left[4-3\left(\frac{t_{\tau}}{t_{sc}}\right)^{-1/2} 
\left(\frac{t}{t_{\tau}}\right)^{-1/4}\right] 
& $(t_{sc} < t_{\tau} < t)$ \cr
A\left(\frac{r}{R}\right)^{-2}\left(\frac{t}{t_{\tau}}\right)^{-1} & $(t_{\tau} 
< t$ and $t_{\tau} \ll t_{sc})$ },
\label{limlc}
\end{equation}
where
\begin{equation}
A = \frac{3c\tau_{\nu (1+z)}}{4R(1+z)}e^{-f\tau_{\nu (1+z)}}F_\nu dt,
\end{equation}
\begin{equation}
t_{sc} = \frac{1}{4}\frac{R(1+z)}{c}\left(\frac{r}{R}\right)
\end{equation}
is the time when the scattering cross section begins to fall off due to 
scattering at greater angles, and 
\begin{equation}
t_{\tau} \approx \frac{3}{\tau_{\nu 
(1+z)}}\frac{R(1+z)}{c}\left(\frac{r}{R}\right)^2
\end{equation}
is the time when the optical depth grows greater by 1 due to the greater path 
length at greater scattering angles.  This interaction of the dust and light 
geometries is depicted in the top panel of Figure 3.  We plot example light 
curves for different values of $f$, $R(1+z)$, and $\tau_{\nu (1+z)}$ in the 
top panels of Figures 4 -- 7; however, consider the following limiting cases:

\noindent{$\bf f=0$ (Figures 4 and 5, top panels):}  If $\tau_{\nu (1+z)} \la 
12$, then $t_{sc} \la t_{\tau}$.  Consequently, $F^E_{\nu}(t) \propto t^0$ 
from $t = 0$ until $t \approx t_{sc}$, after which the light curve slowly 
rolls over to $F^E_{\nu}(t) \propto t^{-3/2}$.  If $\tau_{\nu (1+z)} \la 
1.5$, then this continues until $t \approx t_{off}$, after which 
$F^E_{\nu}(t) \propto t^{-2}$.  If $1.5 \la \tau_{\nu (1+z)} \la 12$, then 
this continues until $t \approx t_{\tau}$, after which the light curve slowly 
rolls over to $F^E_{\nu}(t) \propto t^{-7/4}$ until after $t = t_{off}$.  
After this time, $F^E_{\nu}(t) \propto t^{-2}$.  If $\tau_{\nu (1+z)} \gg 
12$, then $t_{\tau}$ is initially less than $t_{sc}$.  Consequently, 
$F^E_{\nu}(t) \propto t^0$ from $t \approx 0$ until $t \approx t_{\tau}$, 
after which $F^E_{\nu}(t) \propto t^{-1}$ until after $t = t_{off}$.  After 
this time, $F^E_{\nu}(t) \propto t^{-2}$.

\noindent{$\bf f=1$ (Figures 6 and 7, top panels):}  The light curve has the 
same dependence on time as in the $f = 0$ case except that it does not ``turn 
on'' until $\sim t_{on}$.  Also, since the dust is destroyed along the line 
of sight, the pulse of light is a factor of $e^{\tau_{\nu (1+z)}}$ brighter 
than in the $f = 0$ case.  Consequently, the light curve, which is normalized 
to the fluence of the pulse of light, is a factor of $e^{-\tau_{\nu (1+z)}}$ 
fainter than in the $f = 0$ case.

Let $\tau_{\nu(1+z)} \propto [\nu(1+z)]^{\alpha_{\nu(1+z)}}$.  The value of 
$\alpha_{\nu(1+z)}$ is canonically assumed to be 1; however, the true value 
is closer to $\approx 1.6$ at NIR values of $\nu(1+z)$, and levels off to 
$\approx 1$ in the UV (Reichart 2000).  The spectral index of the dust echo 
at $\nu$ as a function of $t$ is then given by:
\begin{eqnarray}
a^E_{\nu}(t) & = & \frac{d\log{F^E_{\nu}(t)}}{d\log{\nu}} \\
& \approx & \cases{a_{\nu} + 
\alpha_{\nu(1+z)}\left[1-\frac{1}{2}\frac{t}{t_{\tau}}-f\tau_{\nu 
(1+z)}\right] & $(t < t_{sc}$ and $t < t_{\tau})$ \cr
a_{\nu} + 
\alpha_{\nu(1+z)}\left[1-\frac{1}{10}\frac{8-3\left(\frac{t}{t_{sc}}\right)^{-5/
2}}{4-3\left(\frac{t}{t_{sc}}\right)^{-1/2}}\frac{t}{t_{\tau}}-f\tau_{\nu 
(1+z)}\right] & $(t_{sc} < t < t_{\tau})$ \cr
a_{\nu} + 
\alpha_{\nu(1+z)} \left[1-\frac{1}{10}\frac{8-3 
\left(\frac{t_{\tau}}{t_{sc}}\right)^{-5/2} 
\left(\frac{t}{t_{\tau}}\right)^{-5/4}}{4-3 
\left(\frac{t_{\tau}}{t_{sc}}\right)^{-1/2} 
\left(\frac{t}{t_{\tau}}\right)^{-1/4}}-f\tau_{\nu (1+z)}\right] 
& $(t_{sc} < t_{\tau} < t)$ \cr
a_{\nu} + \alpha_{\nu(1+z)}\left[\frac{1}{2}-f\tau_{\nu (1+z)}\right] & 
$(t_{\tau} < t$ and $t_{\tau} \ll t_{sc})$ \cr},
\label{limspec}
\end{eqnarray}
where $a_{\nu}$ is the spectral index of the pulse of light.  This 
approximation holds if $f = 0$, or if $f = 1$ and $\theta_{jet} \ll {\rm 
min}\{\theta, {(\theta/\tau)(r/R)}\}$.  We plot how the spectral index of 
the dust echo differs from the spectral index of the pulse of light as a 
function of time for different values of $f$, $R(1+z)$, and $\tau_{\nu 
(1+z)}$ in the bottom panels of Figures 4 -- 7; however, consider the 
following limiting cases:

\noindent{$\bf f=0$ (Figures 4 and 5, bottom panels):}  Since blue light 
scatters with a higher probability than red light, and since both the dust 
echo and the pulse of light are extinguished equally at $t = 0$, the dust 
echo is initially bluer than the pulse of light, independent of the value of 
$\tau_{\nu (1+z)}$:  $a^E_{\nu}(t) - a_{\nu} \approx \alpha_{\nu(1+z)} 
\approx 1$.  If $\tau_{\nu (1+z)} \la 12$, then $t_{sc} \la t_{\tau}$.  
Consequently, the dust echo reddens to $a^E_{\nu}(t) - a_{\nu} \approx 
\alpha_{\nu(1+z)}(1-\tau_{\nu (1+z)}/24)$ at $t \approx t_{sc}$.  If 
$\tau_{\nu (1+z)} \la 1.5$, then it grows redder to $a^E_{\nu}(t) - a_{\nu} 
\approx \alpha_{\nu(1+z)}(1-\tau_{\nu (1+z)}/8)$ at $t \approx t_{off}$, 
after which it grows bluer and returns to its original color.  If $1.5 \la 
\tau_{\nu (1+z)} \la 12$, then it grows redder to $0.5\alpha_{\nu(1+z)} \la 
a^E_{\nu}(t) - a_{\nu} \la 0.7\alpha_{\nu(1+z)}$ at $t \approx t_{\tau}$, 
after which it grows bluer to a color of $a^E_{\nu}(t) - a_{\nu} \approx 
0.8\alpha_{\nu(1+z)}$ until after $t = t_{off}$.  After this time, it grows 
bluer still and returns to its original color.  If $\tau_{\nu (1+z)} \gg 12$, 
then $t_{\tau}$ is initially less than $t_{sc}$.  Consequently, the dust echo 
reddens to $a^E_{\nu}(t) - a_{\nu} \approx 0.5\alpha_{\nu(1+z)}$ at $t 
\approx t_{\tau}$, after which its color does not change until after $t = 
t_{off}$.  After this time, its color grows bluer and returns to its original 
color.

\noindent{$\bf f=1$ (Figures 6 and 7, bottom panels):}  The spectral index of 
the dust echo has the same dependence on time as in the $f = 0$ case, but is 
redder than the pulse of light by an additional $\tau_{\nu (1+z)}$:  since 
the dust is destroyed along the line of sight, the pulse of light is not 
similarly reddened.

In any case, if $\tau_{\nu (1+z)} \la 12$, then the spectral index of the 
dust echo does not change in time by more than $0.5\alpha_{\nu(1+z)} \approx 
0.5$, and if $\tau_{\nu (1+z)} \ll 1$, then it does not change at all.  If 
$\tau_{\nu (1+z)} \ga 12$, then the dust echo is probably too extinguished to 
be observed; consequently, we do not explore this case in detail in this 
paper.

\subsection{Dust Echoes from a Scattered Collimated Pulse of Light}

We now model the more general case of the dust echo of a collimated pulse of 
light that is emitted from a distance that is much less than $R$.  Let the 
pulse of light be emitted between $0 < \theta < \theta_{col}$ and 
$\pi-\theta_{col} < \theta < \pi$ (we assume bipolar emission), where 
$\theta_{col} > \theta_{jet}$.  The only difference from the isotropic case 
is that $dP^{sc}_{\nu (1+z)}(r,\theta) = 0$ if $\theta_{col} < \theta < \pi - 
\theta_{col}$.  Consequently, the light curve is again given by Equation 
(\ref{pulse}), but with
\begin{eqnarray}
P(r') = \cases{0 & $[r' < r(\pi-f\theta_{jet})]$ \cr
1 & $[r(\pi-f\theta_{jet}) < r' < r(\pi-\theta_{col})]$ \cr
0 & $[r(\pi-\theta_{col}) < r' < r(\theta_{col})]$ \cr
1 & $[r(\theta_{col}) < r' < r(f\theta_{jet})]$ \cr
0 & $[r' > r(f\theta_{jet})]$ },
\label{pr}
\end{eqnarray}
\begin{equation}
r = \cases{R & $[t < t(\theta_{col})]$ \cr 
r(\theta_{col}) & $[t(\theta_{col}) < t < t(\pi-\theta_{col})]$ \cr
R & $[t(\pi-\theta_{col}) < t < t(\pi-f\theta_{jet})]$ \cr
r(\pi-f\theta_{jet}) &  $[t > t(\pi-f\theta_{jet})]$},
\label{r}
\end{equation}
and
\begin{equation}
\theta = \cases{\theta(R) & $[t < t(\theta_{col})]$ \cr
\theta_{col} & $[t(\theta_{col}) < t < t(\pi-\theta_{col})]$ \cr
\theta(R) & $[t(\pi-\theta_{col}) < t < t(\pi-f\theta_{jet})]$ \cr
\pi-f\theta_{jet} & $[t > t(\pi-f\theta_{jet})]$}.
\label{theta}
\end{equation}
This interaction of the dust and light geometries is depicted in the bottom 
panel of Figure 3.  

We plot example light curves and color histories for different values of 
$\theta_{col}$ and $\tau_{\nu (1+z)}$ in Figures 8 and 9, but basically, they 
are the same as in the isotropic case until $t \approx t(\theta_{col})$.  
After this time, the light curve falls off as $F^E_{\nu}(t) \propto t^{-2}$ 
if $t < t_{\tau}$, as $F^E_{\nu}(t) \propto t^{-7/4}$ if $t > t_{\tau}$ and 
$t_{sc} < t_{\tau}$ [$\tau_{\nu(1+z)} \la 12r/R$], and as $F^E_{\nu}(t) 
\propto t^{-1}$ if $t > t_{\tau}$ and $t_{\tau} < t_{sc}$ [$\tau_{\nu(1+z)} 
\gg 12r/R$], because the observed light no longer includes light that 
scattered close to the evacuated region where the density (and the optical 
depth) is highest.  A direct consequence of this is that the dust echo grows 
bluer.  The light curve and color history partially recover their isotropic 
characteristics after $t \approx t(\pi-\theta_{col})$, when back-scattered 
light from the far side of the evacuated region can first reach the observer; 
however, the light curve falls off rapidly after $t \approx t_{off}$.

\subsection{Dust Echoes from Scattered Afterglow Light}

We now model the even more general case of the dust echo of collimated light 
that is emitted as a function of time (i.e., an afterglow) from a distance 
that is much less than $R$.  Since the afterglow emits light continuously, at 
any time $t$ after the burst, we receive scattered light that was emitted by 
the afterglow at all prior observer-frame times:  $0 < t' < t$.  
Consequently, the dust echo light curve is given by integrating Equation 
(\ref{pulse}) over these times:
\begin{equation}
F^E_{\nu}(t) =  \frac{3c\tau_{\nu 
(1+z)}}{2R(1+z)}\int_0^t\int_r^{\infty}P(r')\left(\frac{r'}{R}\right)^{-3} 
\frac{e^{-[\tau_{\nu (1+z)}(r',\theta)-(1-f)\tau_{\nu 
(1+z)}]}}{(5-4\cos{\theta})^{3/2}} d\left(\frac{r'}{R}\right)F_\nu(t')dt',
\end{equation}
where $P(r')$, $r$, and $\theta$ are given by Equations (\ref{pr}), 
(\ref{r}), and (\ref{theta}), respectively, but with
\begin{equation}
r(\theta) = \frac{c(t-t')}{(1-\cos{\theta})(1+z)},
\end{equation}
\begin{equation}
\theta(r) = \cos^{-1}\left[1-\frac{c(t-t')}{r(1+z)}\right],
\end{equation}
and
\begin{equation}
t(\theta) = t'+\frac{(1-\cos{\theta})R(1+z)}{c}.
\end{equation}
$F_\nu(t')$ is the light curve of the afterglow, and $\theta_{col}$ is now a 
function of time that we approximate by:
\begin{equation}
\theta_{col} = {\rm min}\left\{(\theta_{jet}^2+\gamma^{-2})^{1/2}, 
\frac{\pi}{2}\right\}.
\end{equation}
Let $t_{jet}$ be the time when $\gamma = \theta_{jet}^{-1}$.  Then from 
Equation (\ref{gamma}), 
\begin{equation}
\gamma = 
\cases{10\left(\frac{\theta_{jet}}{0.1}\right)^{-1} 
\left(\frac{t'}{t_{jet}}\right)^{-1/4} 
& $(t' < t_{jet})$ \cr
10\left(\frac{\theta_{jet}}{0.1}\right)^{-1} 
\left(\frac{t'}{t_{jet}}\right)^{-1/2} 
& $(t' > t_{jet})$}, 
\end{equation}
and
\begin{equation}
t_{jet} = 
5.5(1+z)\left[\frac{n(\rm{1\,pc})}{\rm{1\,cm}^3}\right]^{-1}\left(\frac{E\theta_
{jet}^2}{10^{52}\rm{\,erg}}\right)\left(\frac{\theta_{jet}}{0.1}\right)^2{\rm\,h
r}.
\label{tjet}
\end{equation}
The value of $\gamma \propto (t')^{-1/2}$ after $t_{jet}$ (e.g., Sari, Piran, 
\& Halpern 1999); this is true in either a $n(r) \propto r^0$ or a $n(r) \propto 
r^{-2}$ circumburst medium.  

Finally, we model the light curve of the afterglow.  In this paper, we simply 
take:
\begin{eqnarray}
F_{\nu}^E(t') = \cases{F_{\nu}^E & $(t' < t_0)$ \cr
F_{\nu}^E\left(\frac{t'}{t_0}\right)^{-1.2} & $(t_0 < t' < t_{jet})$ \cr
F_{\nu}^E\left(\frac{t_{jet}}{t_0}\right)^{-1.2}\left(\frac{t'}{t_{jet}}\right)^
{-2.4} & $(t' > t_{jet})$}, 
\label{afterglow}
\end{eqnarray}
which is typical of observed afterglows after $t_0$, where $t_0$ may be as 
little as minutes or as much as hours after the burst (e.g., Sari, Piran, \& 
Narayan 1998).  Before $t_0$, the light curve can depend on time in a variety 
of ways (e.g., Sari, Piran, \& Narayan 1998); however, we do not model it any 
more carefully than a constant.  Instead, we show that the value of $t_0$ has 
little effect on the light curve and color history of the dust echo (see Figure 
13).  

We plot example light curves and color histories for different values of 
$\theta_{jet}$, $t_{jet}$, and $t_0$ in Figures 10 -- 13.  The results are 
similar to the case of a collimated pulse of light with collimation 
half-angle given by the value of $\theta_{col}$ when most of the afterglow 
light is emitted; i.e., by the value of $\theta_{jet}$, since most of the 
afterglow light is emitted between $t_0 < t' < t_{jet}$, and $\theta_{col} 
\approx \theta_{jet}$ during these times.

In Figure 10, we take $f = 1$, $R(1+z) = 0.1$ pc, $\tau_{\nu (1+z)} = 1$, $t_0 = 
1$ hr, 
$\theta_{jet} = 5^{\circ}$, $10^{\circ}$, $20^{\circ}$, and $40^{\circ}$, and 
$t_{jet} = 0.35$, 1.4, 5.6, and 22 dy, respectively.  These values of 
$t_{jet}$ are given by Equation (\ref{tjet}) with $z = 1$, $n($1 pc$) = 1$ 
cm$^{-3}$, $E\theta_{jet}^2 = 10^{52}$ erg, and these values of 
$\theta_{jet}$.  Clearly, for these values of the parameters, the dust echo 
of the afterglow is substantially fainter than the afterglow itself.  
Changing the value of $\tau_{\nu (1+z)}$ only makes the dust echo fainter 
relative to the afterglow:  decreasing $\tau_{\nu (1+z)}$ decreases the 
amount of light scattered into the line of sight, and increasing $\tau_{\nu 
(1+z)}$ increases how much the dust echo is extinguished, but does not 
increase how much the afterglow is extinguished since $f = 1$.   The dust 
echo can be made brighter than the afterglow if $f = 0$ and $\tau_{\nu 
(1+z)}$ is substantially increased, but in this scenario, both the afterglow 
and the dust echo are substantially extinguished and consequently difficult 
to detect.  This leaves only three scenarios in which a dust echo can be 
brighter than an afterglow and detectable:  (1) the value of $R(1+z)$ is 
substantially greater than 0.1 pc, in which case the dust echo peaks at a 
substantially later time (but with a substantially lower peak brightness), 
after the light curve of the afterglow has rolled over and ``gotten out of 
the way" of the dust echo (Figure 11);  (2) the value of $(1+z)[n($1 
pc$)]^{-1}E\theta_{jet}^2$ is substantially less than the value assumed 
above, in which case the light curve of the afterglow rolls over at a 
substantially earlier time, again ``getting out of the way" of the dust echo 
(Figure 12); and (3) the dust echo is dominated by scattered light not from 
the afterglow, but rather, from the optical flash and/or optical light from 
the burst itself.  This last scenario can be modeled simply as a collimated 
pulse of light (\S 2.2) with $\theta_{col} \ga \theta_{jet}$.  We show that the 
choice of $t_0$ has little effect on the light curve and color history of the 
dust echo of the afterglow in Figure 13.

One aspect of the dust and light geometry that we do not model in this paper 
is that the afterglow is not emitted from the center of the evacuated region, 
but rather from the head of the jet, which is moving away from the center 
(toward the observer) at relativistic speeds.  Since $r \approx 
4\gamma^2ct/(1+z)$ (\S 1), the head of the jet reaches the radius $r$ at 
which most of the scattering is occurring at time 
\begin{equation}
t_{r} = 
8.9(1+z)^2\left(\frac{\theta_{jet}}{0.1}\right)^4\left(\frac{t_{jet}}{1{\rm\,dy}
}\right)^{-1}\left(\frac{r}{1{\rm\,pc}}\right)^2{\rm\,dy}.
\end{equation}
Our model holds as long as most of the light is emitted before some 
reasonable fraction of $t_r$.  For afterglow light curves like the one 
described in Equation (\ref{afterglow}), most of the afterglow light is 
emitted before $t_{jet}$, by a factor of a few.  Consequently, our model only 
holds as long as $t_r \ga t_{jet}$, or
\begin{equation}
r \ga 
0.08\left[\frac{n(\rm{1\,pc})}{\rm{1\,cm}^3}\right]^{-1}\left(\frac{E\theta_{jet
}^2}{10^{52}\rm{\,erg}}\right){\rm\,pc}.
\end{equation}
However, if the dust echo is dominated by scattered light from the optical 
flash and/or optical light from the burst, as opposed to light from the 
afterglow, then the light is obviously emitted long before $t_r$, and 
consequently, the model holds for substantially smaller values of $r$ as well. 

\section{Dust Echoes from Absorbed and Thermally Re-Emitted Light}

The dust echo from absorbed and thermally re-emitted light is easier to model 
than the dust echo from scattered light:  unlike Equation (\ref{scat}), the 
differential cross section for dust absorption has no dependence on $\theta$: 
\begin{equation}
\frac{d\sigma_{\nu (1+z)}^{abs}}{d\Omega} = \frac{(1-\epsilon_{\nu 
(1+z)})\sigma_{\nu (1+z)}^{tot}}{4\pi}.
\end{equation}
Consequently, the light curve of the dust echo of the afterglow is given by 
\begin{equation}
F^E_{IR}(t) =  \frac{c\tau_{\nu 
(1+z)}}{R(1+z)}\int_0^t\int_r^{\infty}P(r')\left(\frac{r'}{R}\right)^{-3} 
e^{-[\tau_{\nu (1+z)}(r',\theta)-(1-f)\tau_{\nu 
(1+z)}]}d\left(\frac{r'}{R}\right)F_{UV}(t')dt',
\end{equation}
where $F_{IR}(t)$ is the bolometric light curve of the dust echo, and 
$F_{UV}(t')$ is the light curve of the afterglow, integrated over UV values 
of $\nu(1+z)$.  If the re-radiating dust particles had a common temperature 
$T$, the spectrum of the dust echo would be thermal.  However, the 
temperature of the dust particles decreases with radius as $T \propto 
r^{-1/2}$ (Waxman \& Draine 2000); consequently, the spectrum should fall off 
more slowly than a blackbody redward of the spectral peak.  Since dust is 
rapidly sublimated at temperatures $T \ga 2300$ K, the spectral peak should 
occur only at frequencies $\nu \la 1.35\times10^{14}(1+z)^{-1}$ Hz.  For a burst 
at $z = 0$, this is in the K band. 

We plot example light curves for different values of $\theta_{jet}$ in Figure 
14.  However, consider the limiting cases for an isotropic 
pulse of light.  The thermal dust echo has the same limiting cases as the 
scattered dust echo (e.g., Equation \ref{limlc}), except that $t_{sc} = 
\infty$.  Consequently, if $t_{\tau} > t(\theta_{col})$, then $F_{IR}(t) 
\propto t^{0}$ until $t \approx t(\theta_{col})$, after which $F_{IR}(t) 
\propto t^{-2}$ until $t \approx t_{\tau}$.  After this time, $F_{IR}(t) 
\propto t^{-1}$.  If $t_{\tau} < t(\theta_{col})$, then $F_{IR}(t) \propto 
t^0$ until $t \approx t_{\tau}$, after which $F_{IR}(t) \propto t^{-1}$.

\section{Can the Late-Time Afterglow of GRB 970228 Be Explained by a Dust 
Echo?}

No.  First of all, the late-time afterglow of GRB 970228 cannot be explained 
by a thermal dust echo, because its spectral flux distribution (Figure 1) 
peaks between the V and B bands in the source frame, not in or beyond the K 
band (\S 3).  Secondly, the late-time afterglow of GRB 970228 cannot be 
explained by a dust echo due to scattered afterglow light, because its light 
curve (e.g., Reichart 1999) does not roll over at sufficiently early times 
for the dust echo to be able to outshine the afterglow when the second 
component to the light curve peaks, about 20 -- 30 days after the burst (\S 
2.3).  This leaves only the possibility of a dust echo due to scattered light 
from the optical flash and/or optical light from the burst itself.  To 
investigate this possibility, we have fitted the spectral flux distribution 
of Figure 1 with two power laws, one to the reddest three points and one to the 
bluest three points.  We find that the 
spectral index redward of and including the source-frame V band is $a^E_{rV} 
=2.29^{+1.43}_{-0.58}$, and that the spectral index blueward of and including 
the 
source-frame B band is $a^E_{bB} = -3.09^{+0.54}_{-0.55}$.  The corresponding 
optical 
depths as a function of the spectral index of the pre-scattered, 
unextinguished light are given by Equation (\ref{limspec}) with $f$ set to 1.  
Taking $\alpha_{rV} = 1.6$ and $\alpha_{bB} = 1$ (\S 2.1), we find that for 
an intrinsic spectral index of -1/2, $\tau_{rV} = -0.99^{+0.44}_{-0.93}$ and 
$\tau_{bB} 
= 3.34^{+0.60}_{-0.60}$, and for an intrinsic spectral index of 1/3, $\tau_{rV} 
\approx 
-0.47^{+0.44}_{-0.93}$ and $\tau_{bB} = 4.17^{+0.60}_{-0.60}$.  Intrinsic 
spectral indexes of -1/2 and 
1/3 are typical of what one expects for the optical flash blueward and 
redward of the synchrotron peak, respectively, which one expects to be at 
optical/NIR wavelengths during the optical flash (e.g., Sari \& Piran 1999).  
In the case of optical light from the burst itself, one expects a spectral 
index of 1/3 across the optical and NIR bands (e.g., Sari \& Piran 1999).  
Consequently, in either interpretation, the value of $\tau_{rV}$ is 
borderline consistent with being physical (i.e., non-negative), but the rapid 
transition between $A_R$ or $A_V \approx 0$ to $A_B$ or $A_U \approx 3.6\pm0.7$ 
or $4.5\pm0.7$ is 
unphysical.  Consequently, we find that the late-time afterglow of GRB 970228 
cannot be explained by a dust echo.  

Unfortunately, the late-time afterglow of GRB 980326 was not sufficiently 
well sampled to distinguish between a supernova and a dust echo explanation

\section{Discussion and Conclusions}

We have modeled and computed dust echo light curves and spectra for dust and
light geometries that are more realistic than had previously been considered.
In particular, bursts are very likely to be collimated (on both theory and
energetics grounds), and dust may be destroyed by the progenitor, as well as the
burst, optical flash, and/or afterglow.  Also, the optical flash and/or optical
light from the burst itself may result in dust echoes.  Consequently, dust
echoes may be a way to measure the brightness of the optical flash and/or the
optical brightness of the burst.

We find that dust echo spectra can be used to distinguish different types of 
dust echoes from each other, and from supernovae.  For example, thermal dust 
echoes peak in the source-frame K band, while supernovae peak in the 
source-frame V or B bands.  In the absence of extinction (e.g., in the 
source-frame NIR), dust echoes due to scattered light are bluer than the 
pre-scattered light by about 1 -- 1.6 in the spectral index, and 
consequently, can peak in the source-frame V or B bands, or other nearby 
bands, depending on the amount of extinction.  However, dust echoes due to 
scattered light do not peak as sharply as do thermal dust echoes or 
supernovae.  It is for these reasons that we find that the late-time 
afterglow of GRB 970228 cannot be explained by a dust echo.  

Furthermore, dust echo light curves can often be used to distinguish dust 
echoes from supernovae.  While dust echoes can peak about 20(1+z) days after 
a burst, they also can peak substantially earlier and substantially later.  
However, supernovae can peak only at this time.  

Finally, dust echoes can be used to probe the circumburst environment, and 
the effects of the progenitor, burst, optical flash, and/or afterglow on this 
environment.  For example, if the dust echo light curve brightens, levels off 
for a long time, and then fades, then $f = 0$ and dust had not (yet) been 
destroyed along the line of sight.  If the dust echo light curve brightens 
and fades without leveling off, then $f = 1$ and dust had been destroyed 
along the line of sight.  The time at which the dust echo light curve begins 
to fade gives you (most likely) a degenerate combination of $R$ and 
$\theta_{jet}$.  The color of the dust echo, particularly near peak, gives 
you $\tau_{\nu (1+z)}$, and consequently a degenerate combination of $R$ and 
$n($1 pc$)$.  The time at which the afterglow light curve rolls over due to 
lateral expansion of the jet gives you a degenerate combination of 
$\theta_{jet}$ and $n($1 pc$)$.  Consequently, $R$, $\theta_{jet}$, and $n($1 
pc$)$ can be separated and solved for.  Furthermore, the brightness of the 
dust echo, particularly near peak, gives you a degenerate combination of $R$, 
$\tau_{\nu (1+z)}$, and the fluence of the pre-scattered, unextinguished 
light; consequently, this fluence can be solved for as well.  Also, by 
comparing $\tau_{\nu (1+z)}$ to the same quantity measured from the afterglow, 
and to absorption measured in X rays, one can further probe dust destruction 
along the line of sight, and the metals-to-dust ratio off the line of sight 
where dust is not destroyed, respectively.

\acknowledgements
Support for this work was provided by NASA through the Hubble Fellowship grant 
\#HST-SF-01133.01-A from the Space Telescope Science Institute, which is 
operated by the Association of Universities for Research in Astronomy, Inc., 
under NASA contract NAS5-26555.  I am very grateful to Re'em Sari for invaluable 
discussions about the dust and light geometries.  I am also grateful to Don 
Lamb for discussions about dust echoes in general.

\clearpage

\figcaption[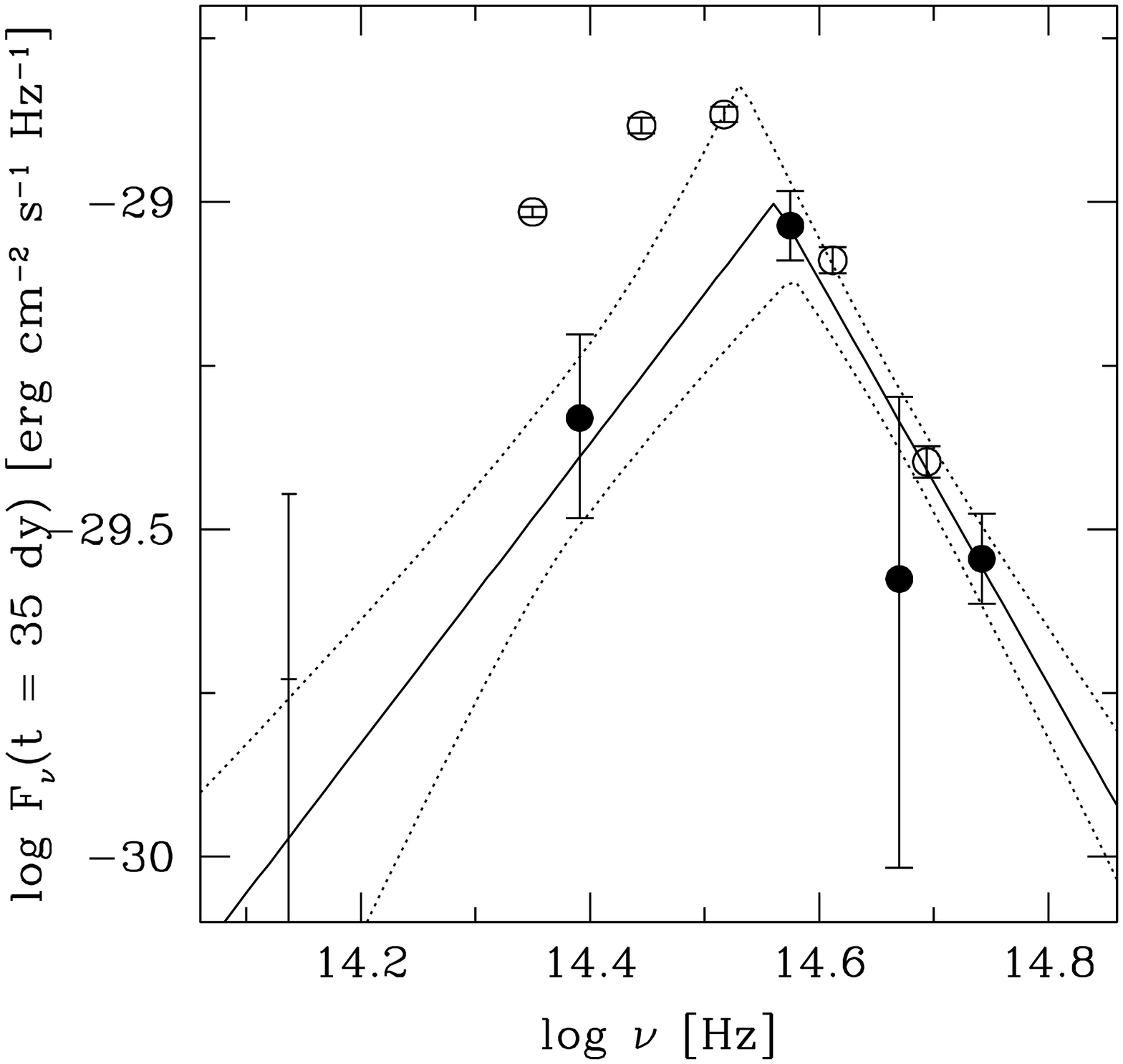]{The K- through V-band spectral flux distribution of the 
late-time afterglow of GRB 970228 after subtracting out the contribution of the 
host galaxy and correcting for Galactic extinction (filled circles and upper 
limit), and the I- through U-band spectral flux distribution of SN 1998bw after 
transforming to the redshift of GRB 970228, $z = 0.695$ (Bloom, Djorgovski, \& 
Kulkarni 2001), and correcting for Galactic extinction (unfilled circles; 
Reichart, Castander, \& Lamb 2000; Reichart, Lamb, \& Castander 2000).  The 
K-band upper limits are 2 and 3 $\sigma$.  We have fitted power laws to the K-, 
J-, and I-band observations, and the I-, R-, and V-band observations of the 
late-time afterglow of GRB 970228 (see \S 4).  The solid lines are the best 
fits, and the dotted lines are the 1 $\sigma$ uncertainties in these fits.  
Clearly, the spectral flux distribution of the late-time afterglow of GRB 970228 
is very blue redward of the peak, and very red blueward of the peak.}

\figcaption[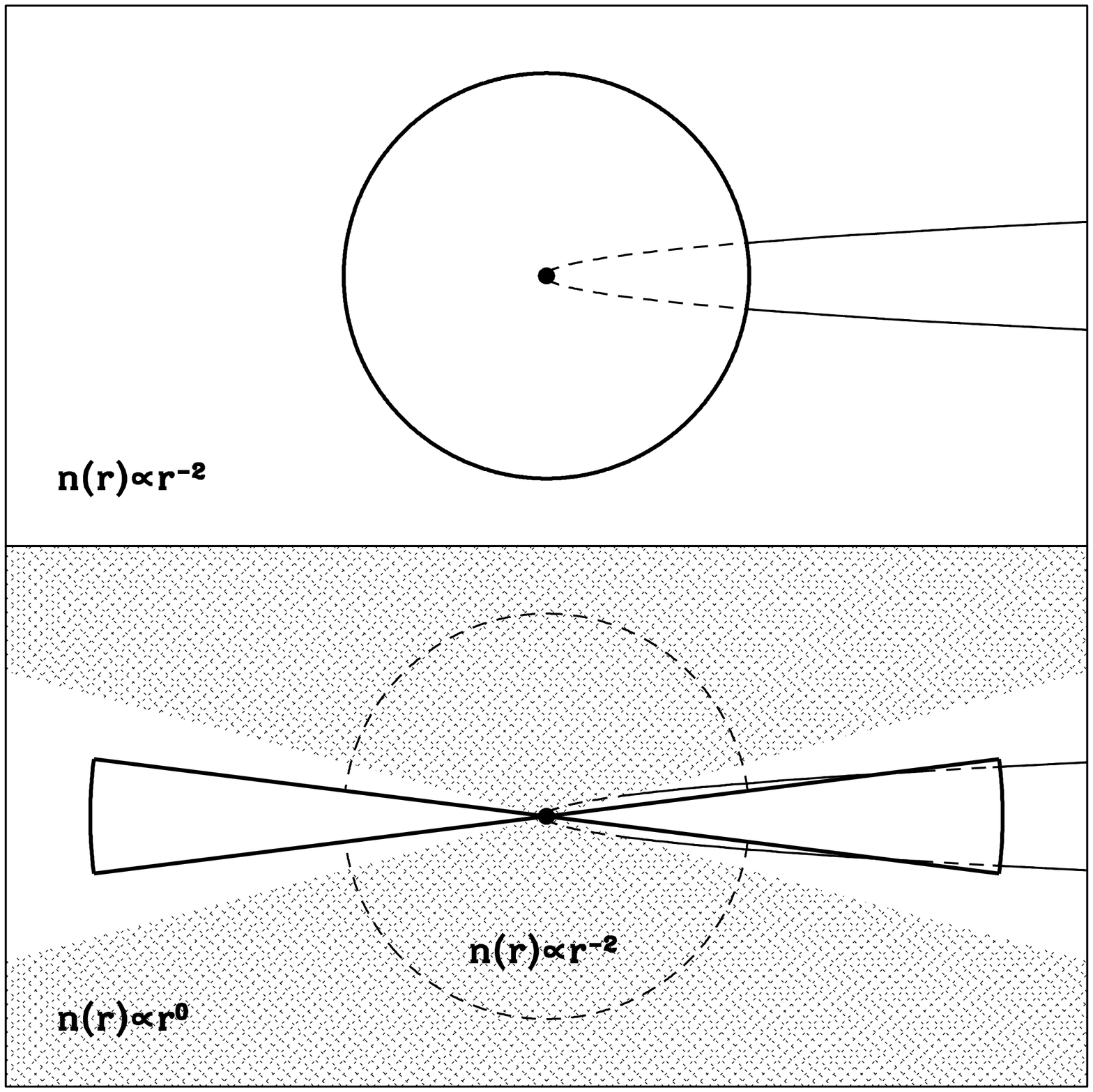]{Dust and light geometries.  Top panel:  Isotropic dust 
destruction and an isotropic pulse of light.  Dust is destroyed isotropically 
about the progenitor, and interior to the thick solid circle, the radius of 
which is $R < R_w$, where $R_w$ is the radius of the wind termination shock.  
Exterior to $R$, $n(r) \propto r^{-2}$ out to $R_w$.  The parabola marks the 
surface on which light must either scatter or be absorbed and thermally 
re-emitted to arrive at an observer to the far right at a constant time $t$ 
after the burst, where $t$ is given by Equation (\ref{delay}).  Light cannot 
scatter or be absorbed and thermally re-emitted on the dotted portion of this 
surface of revolution because dust has been destroyed here.  Bottom panel:  
Collimated dust destruction and a collimated pulse of light.  Dust is destroyed 
within the bipolar cone (thick solid curve) out to a radius $R > R_w$.  The 
radius $R_w$ is marked with a thin dashed circle.  Interior to $R_w$, $n(r) 
\propto r^{-2}$; exterior to $R_w$, $n(r) \propto r^0$.  The shaded area marks 
the region not illuminated by the collimated pulse of light.  The parabola again 
marks a ``surface of constant time".  Light cannot scatter or be absorbed and 
thermally re-emitted on the dotted portions of this surface because either dust 
has been destroyed here, or light does not reach here.  In this dust and light 
geometry, light clearly scatters and is absorbed and thermally re-emitted at 
radii $r \ll R$.}

\figcaption[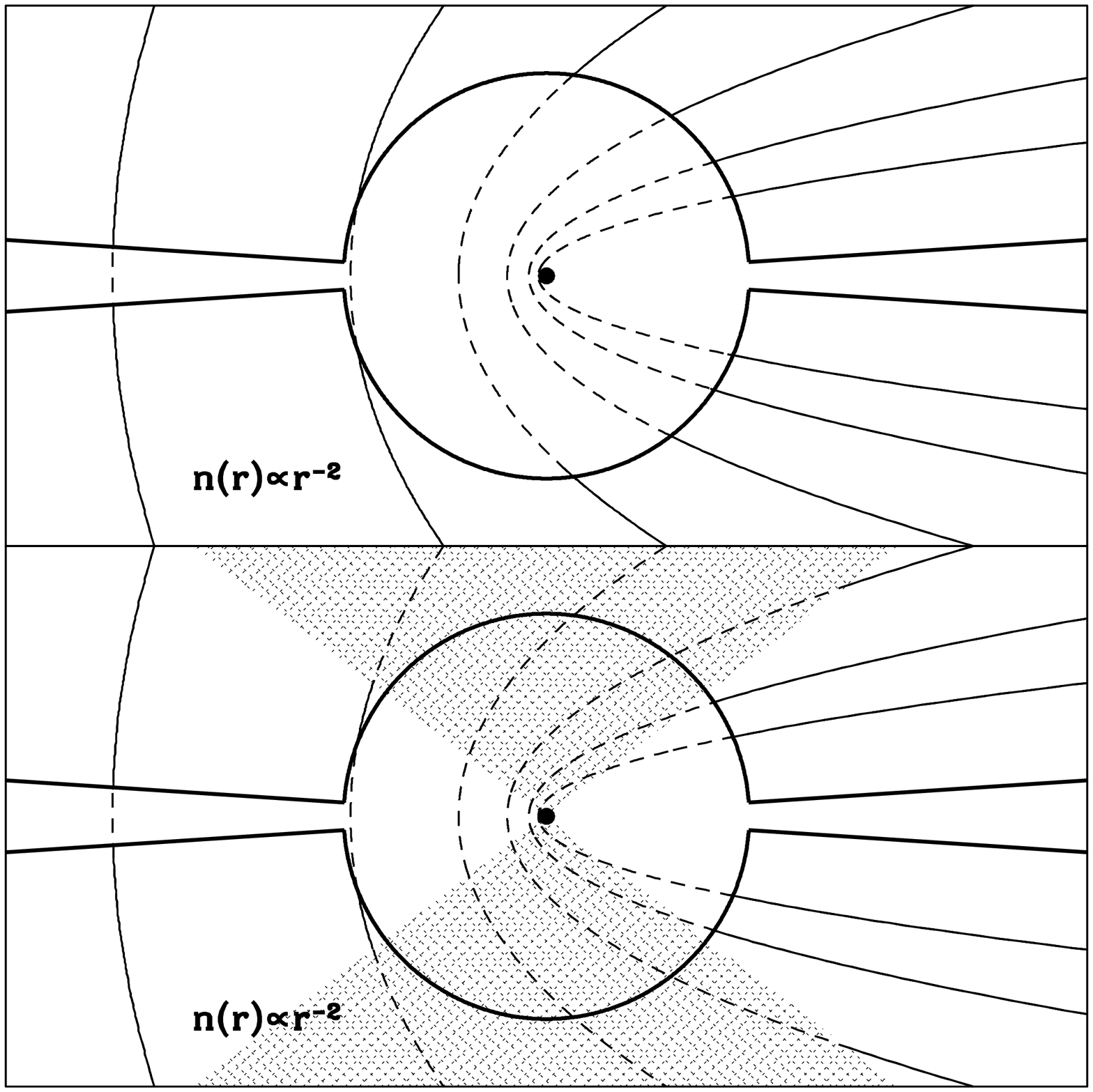]{In both panels, the dust geometry is the same:  dust is 
destroyed interior to a sphere of radius $R < R_w$ by the progenitor, and 
interior to a bipolar cone by the burst, optical flash, and/or afterglow (thick 
solid curve).  Exterior to the thick solid curve, $n(r) \propto r^{-2}$ out to 
$R_w$.  In the top panel, the pulse of light is isotropic; in the bottom panel, 
the pulse of light is collimated.  The parabolas mark surfaces of constant time 
that are separated by factors of two in time.  Clearly, these dust and light 
geometries would result in dust echoes with similar early-time and late-time 
behaviors, but different intermediate-time behaviors.}

\figcaption[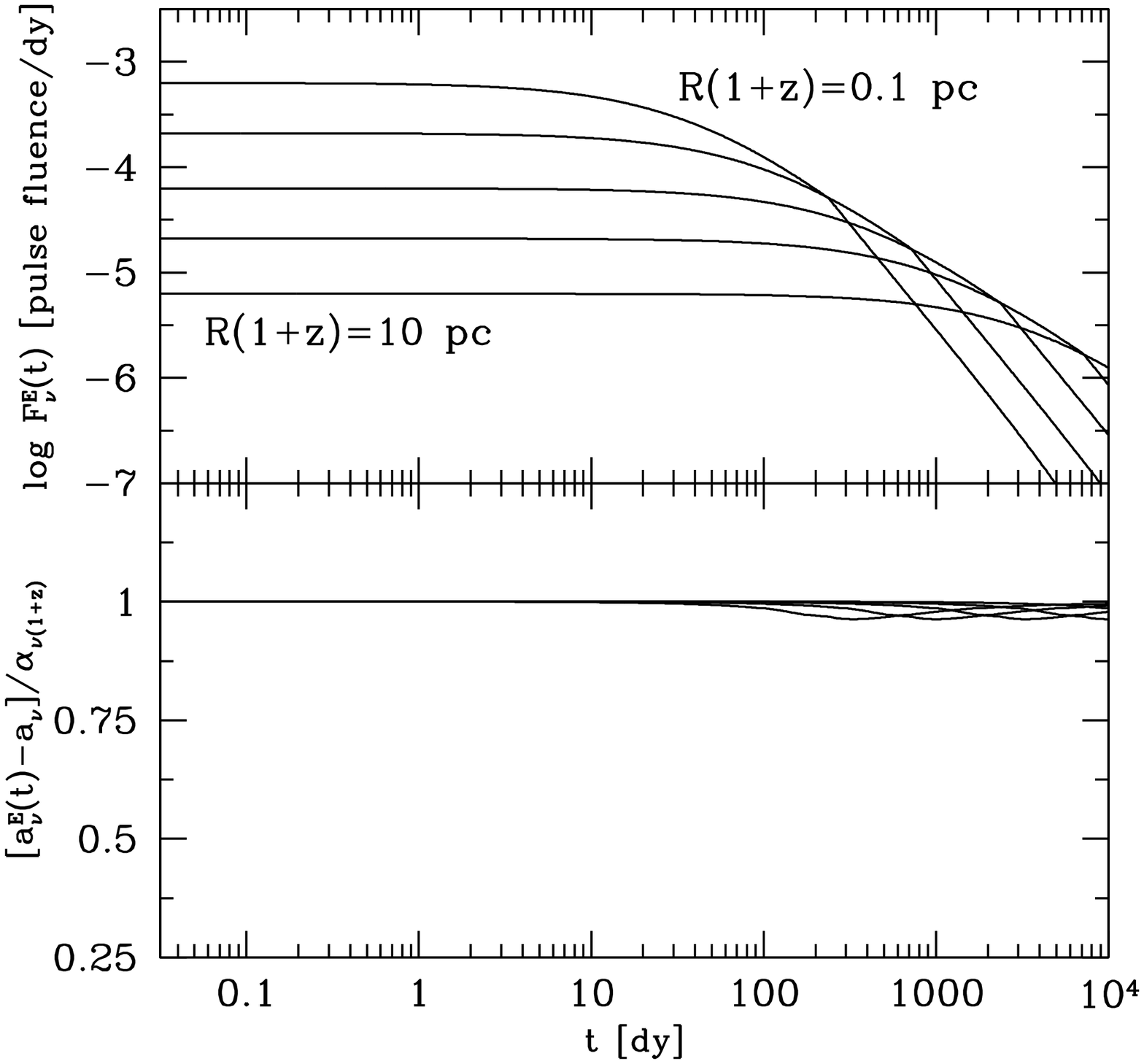]{Light curves (top panel) and color histories (bottom 
panel) of scattering dust echoes of an isotropic pulse of light, with $f = 0$ 
(dust is destroyed isotropically by the progenitor, but not additionally by the 
burst, optical flash, and/or afterglow along the line of sight), $R(1+z) = 0.1$, 
0.3, 1, 3, and 10 pc, and $\tau_{\nu (1+z)} = 0.1$.  Since $f = 0$, the dust 
echo is bluer than the pulse of light.  Since $\tau_{\nu (1+z)} \ll 1$, the 
color of the dust echo is relatively constant with time.}

\figcaption[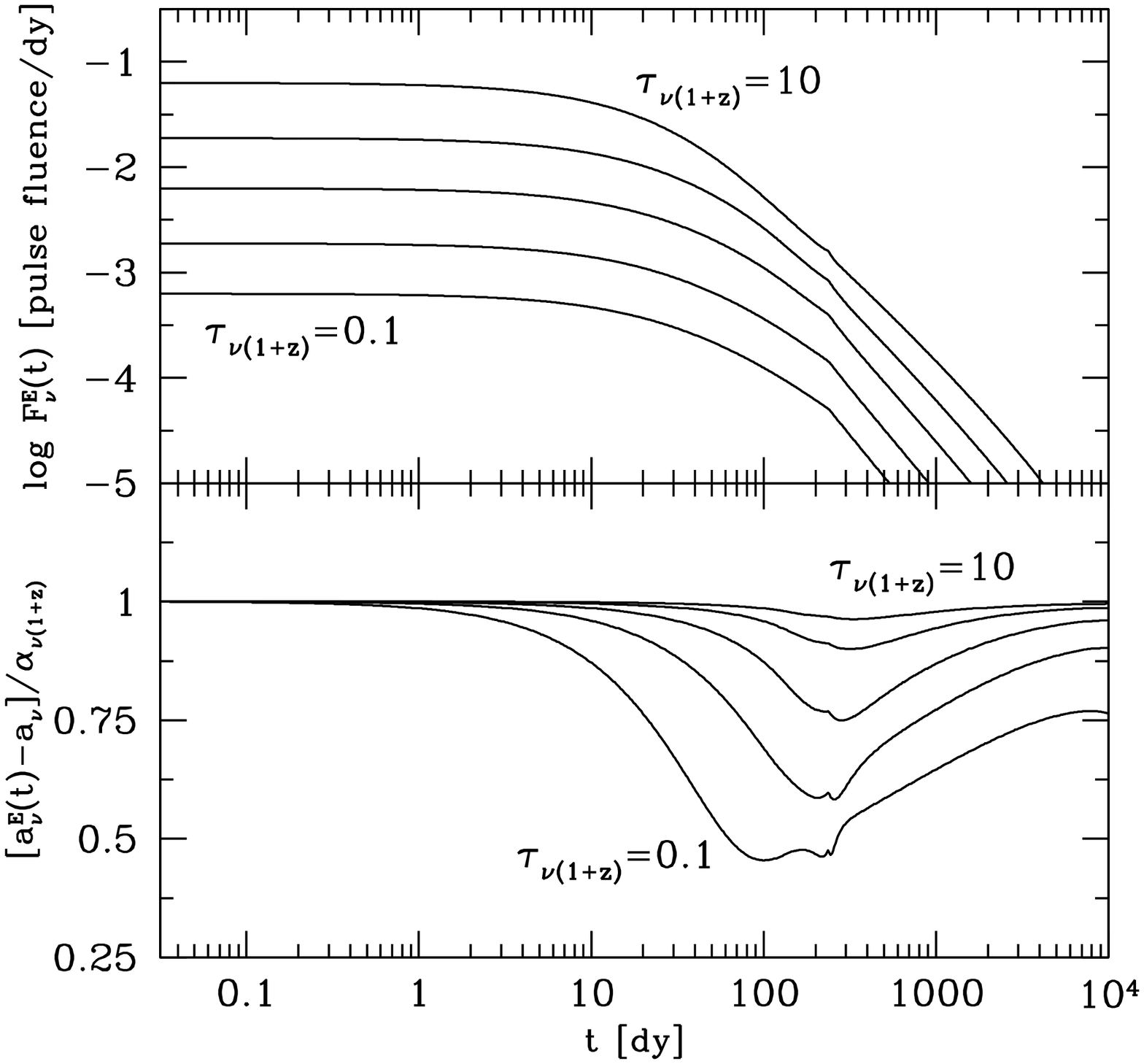]{Light curves (top panel) and color histories (bottom 
panel) of scattering dust echoes of an isotropic pulse of light, with $f = 0$, 
$R(1+z) = 0.1$ pc, and $\tau_{\nu (1+z)} = 0.1$, 0.3, 1, 3, 10.  Since $f = 0$, 
the dust echo is bluer than the pulse of light.  However, when $\tau_{\nu (1+z)} 
\ga 1$, the dust echo does grow somewhat redder when the dust echo begins to 
fade away.  This is due to greater extinction, due to the greater path length 
that the light must travel through the dust at the larger scattering angles.  
These larger scattering angles come into play only at late times, and are in 
fact responsible for the onset of the fading.}

\figcaption[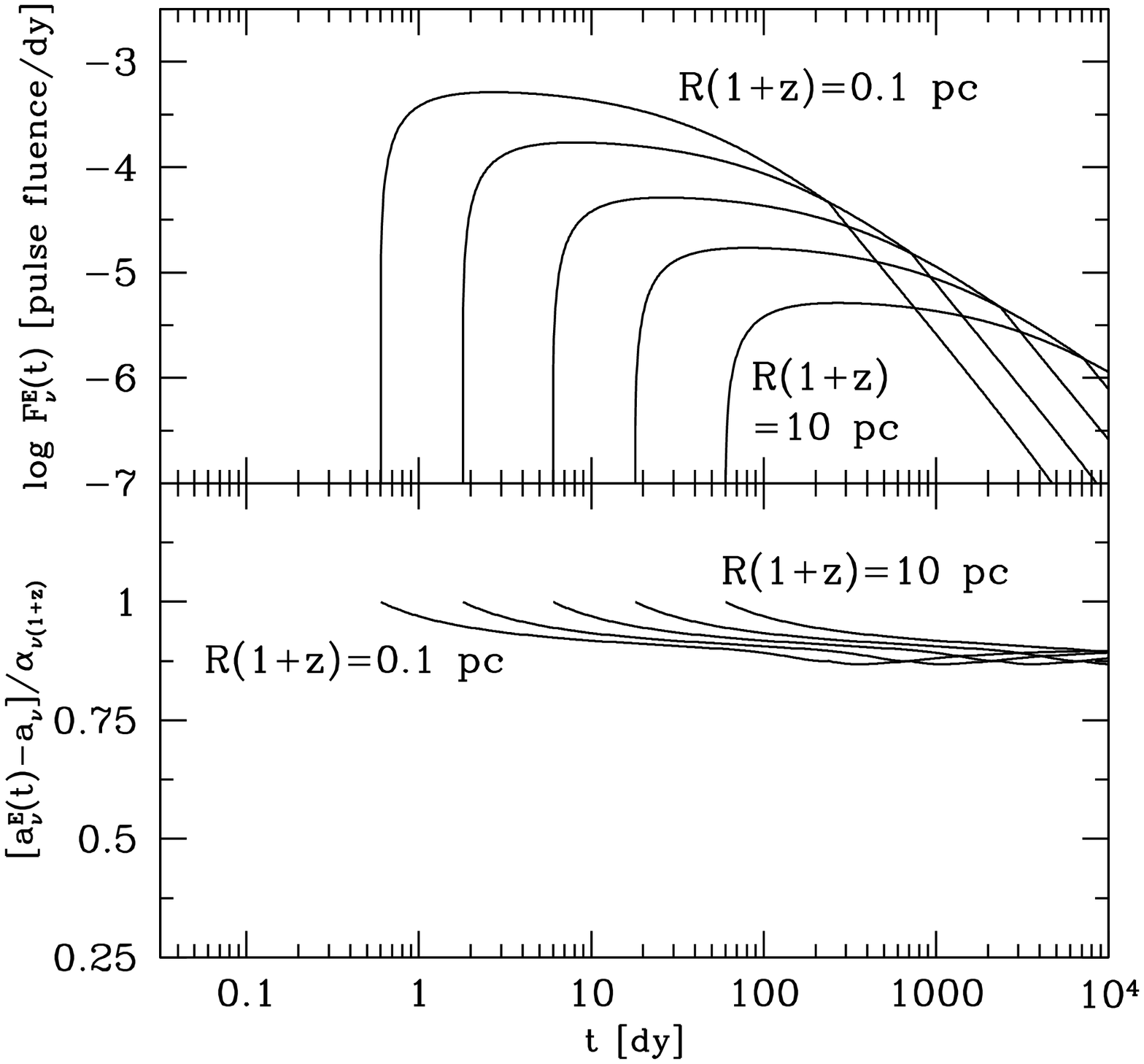]{Light curves (top panel) and color histories (bottom 
panel) of scattering dust echoes of an isotropic pulse of light, with $f = 1$ 
(dust is destroyed both isotropically out to $R$ by the progenitor, and within 
$\theta_{jet}$ of the line of sight out to $r >$ a few times $R$ by the burst, 
optical flash, and/or afterglow), $R(1+z) = 0.1$, 0.3, 1, 3, and 10 pc, 
$\tau_{\nu (1+z)} = 0.1$, and $\theta_{jet} = 0.1$.  The light curves ``turn on" 
once $\theta > \theta_{jet}$.  Since $\tau_{\nu (1+z)} \ll 1$, the color of the 
dust echo is relatively constant with time.}

\figcaption[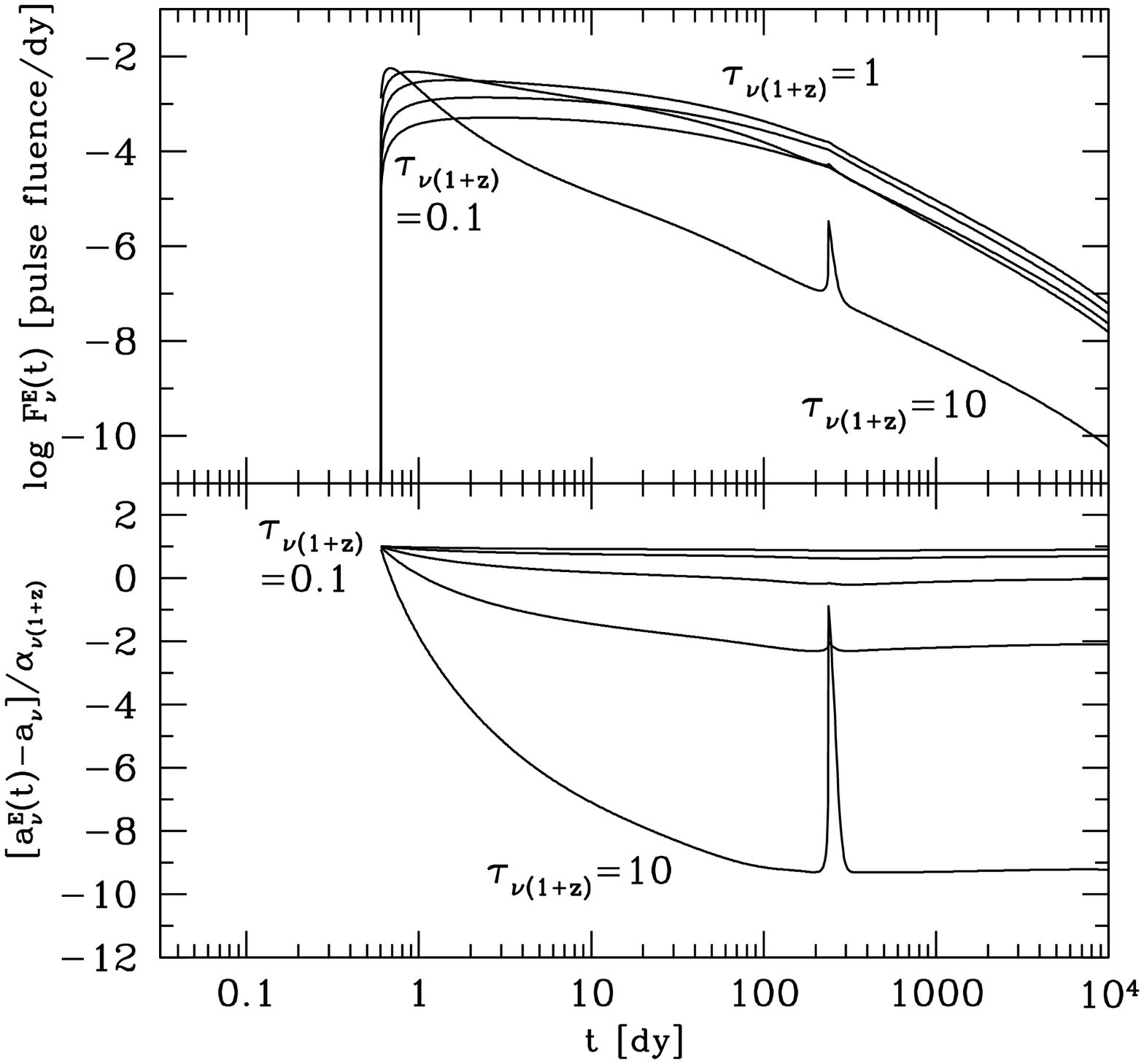]{Light curves (top panel) and color histories (bottom 
panel) of scattering dust echoes of an isotropic pulse of light, with $f = 1$, 
$R(1+z) = 0.1$ pc, $\tau_{\nu (1+z)} = 0.1$, 0.3, 1, 3, 10, and $\theta_{jet} = 
0.1$.  When $\tau_{\nu (1+z)} > 1$, the dust echo is generally fainter and 
redder relative to the pulse of light, because the dust echo is extinguished, 
but the pulse of light, which travels along the dust-evacuated line of sight, is 
not.  However, regardless of the value of $\tau_{\nu (1+z)}$, the dust echo is 
bright initially, when $\theta$ is only slightly greater than $\theta_{jet}$, 
and again when $\theta$ is only slightly less than $\pi - \theta_{jet}$, because 
at these angles, scattered light travels through a minimal amount of dust before 
entering the dust-evacuated cone (of half-angle $\theta_{jet}$) along the line 
of sight.  For the same reason, the dust echo grows significantly bluer at these 
times.}

\figcaption[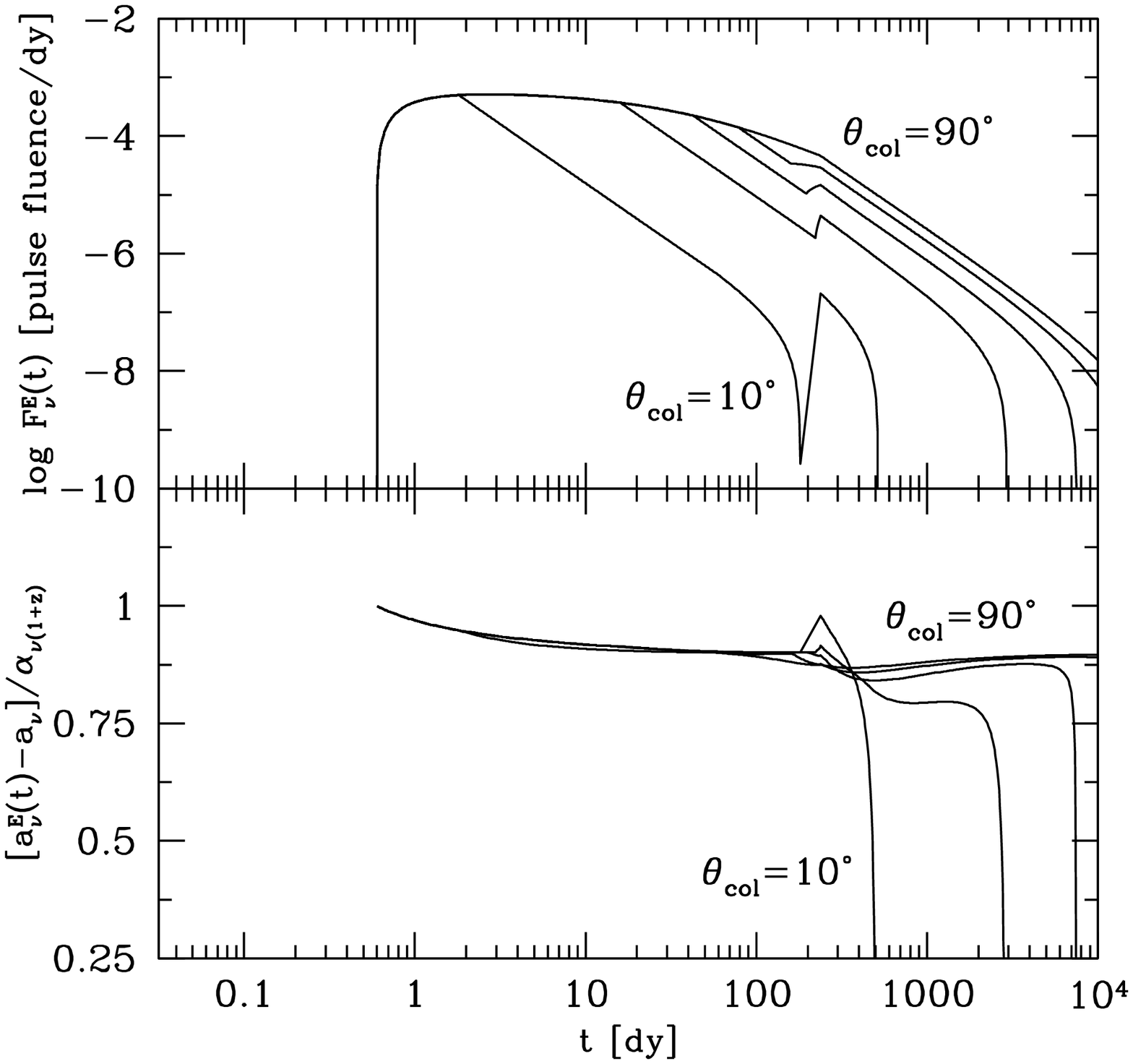]{Light curves (top panel) and color histories (bottom 
panel) of scattering dust echoes of a collimated pulse of light, with $f = 1$, 
$R(1+z) = 0.1$ pc, $\tau_{\nu (1+z)} = 0.1$, $\theta_{jet} = 0.1$ 
($5.7^{\circ}$), and $\theta_{col} = 10^{\circ}$, $30^{\circ}$, $50^{\circ}$, 
$70^{\circ}$, and $90^{\circ}$.  The $\theta_{col} = 90^{\circ}$ light curve and 
color history is the same as in Figure 7.  The other light curves and color 
histories are the same as this one, except that the light that scattered at 
$\theta > \theta_{col}$ no longer contributes.}

\figcaption[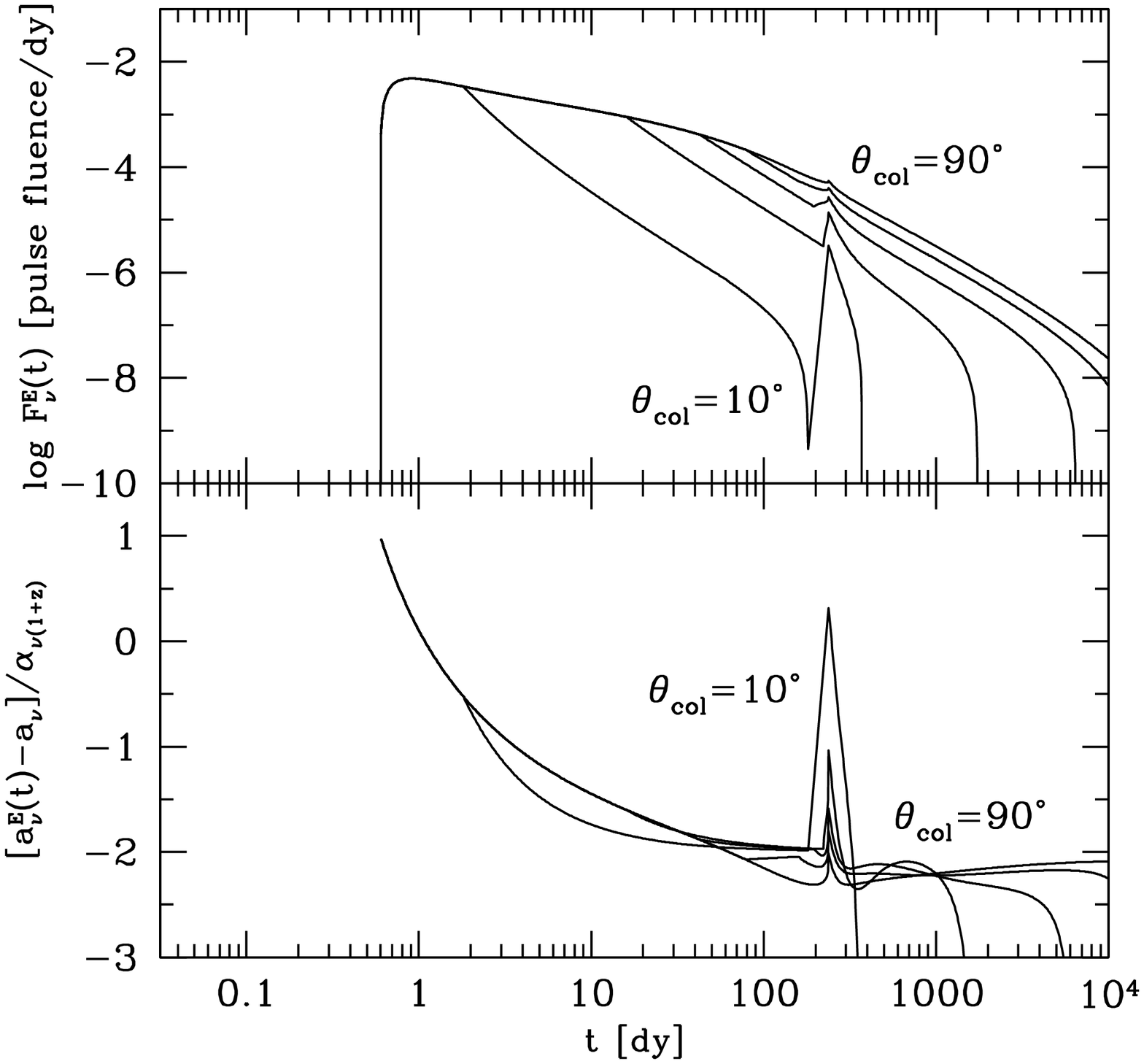]{Same as Figure 8, but with $\tau_{\nu (1+z)} = 3$.}

\figcaption[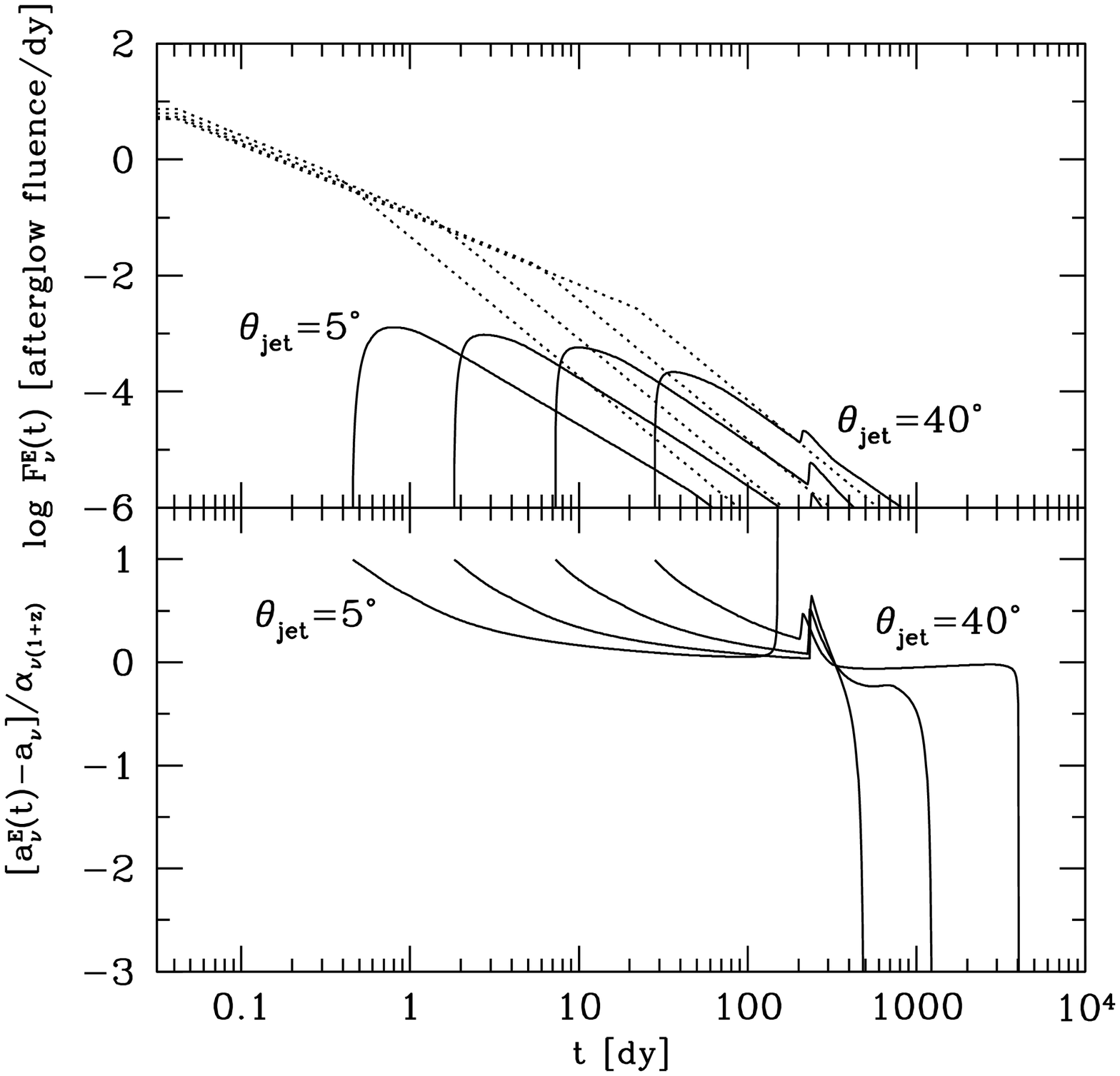]{Light curves (top panel) and color histories (bottom 
panel) of scattering dust echoes (solid lines) of afterglow light (dotted 
lines), with $f = 1$, $R(1+z) = 0.1$ pc, $\tau_{\nu (1+z)} = 1$, $t_0 = 1$ hr, 
$\theta_{jet} = 5^{\circ}$, $10^{\circ}$, $20^{\circ}$, and $40^{\circ}$, and 
$[(1+z)/2][n(\rm{1\,pc})/\rm{1\,cm}^3]^{-1}(E\theta_{jet}^2/10^{52}\rm{\,erg}) = 
1$.  The corresponding values of $t_{jet}$ are 0.35, 1.4, 5.6, and 22 dy, 
respectively (Equation \ref{tjet}).  Clearly, the dust echoes of these 
afterglows would be lost in the brightness of the afterglow.}

\figcaption[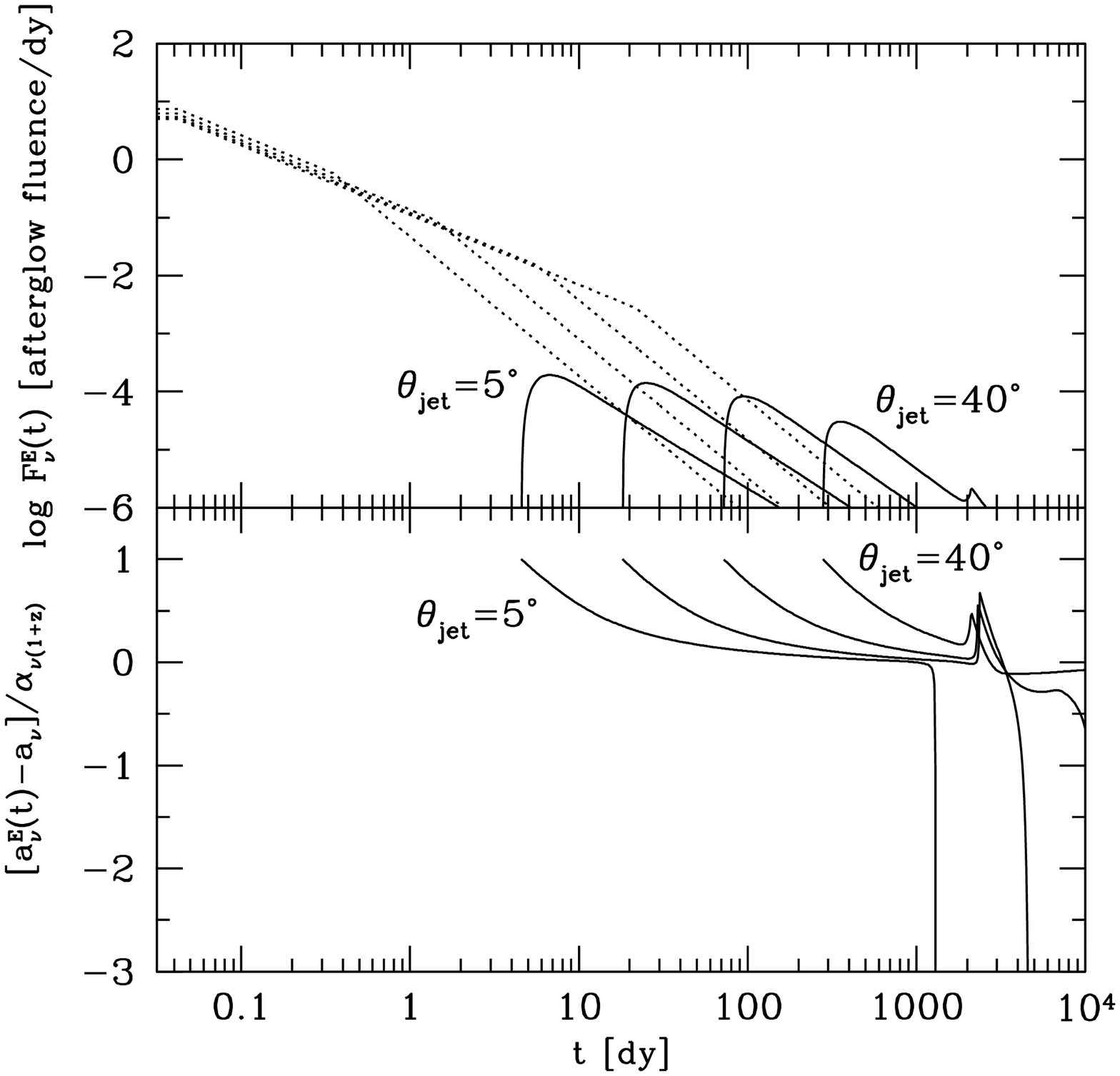]{Same as Figure 10, but with $R(1+z) = 1$ pc.  The dust 
echoes are brighter than the afterglows at times substantially later than 
$t_{jet}$.}

\figcaption[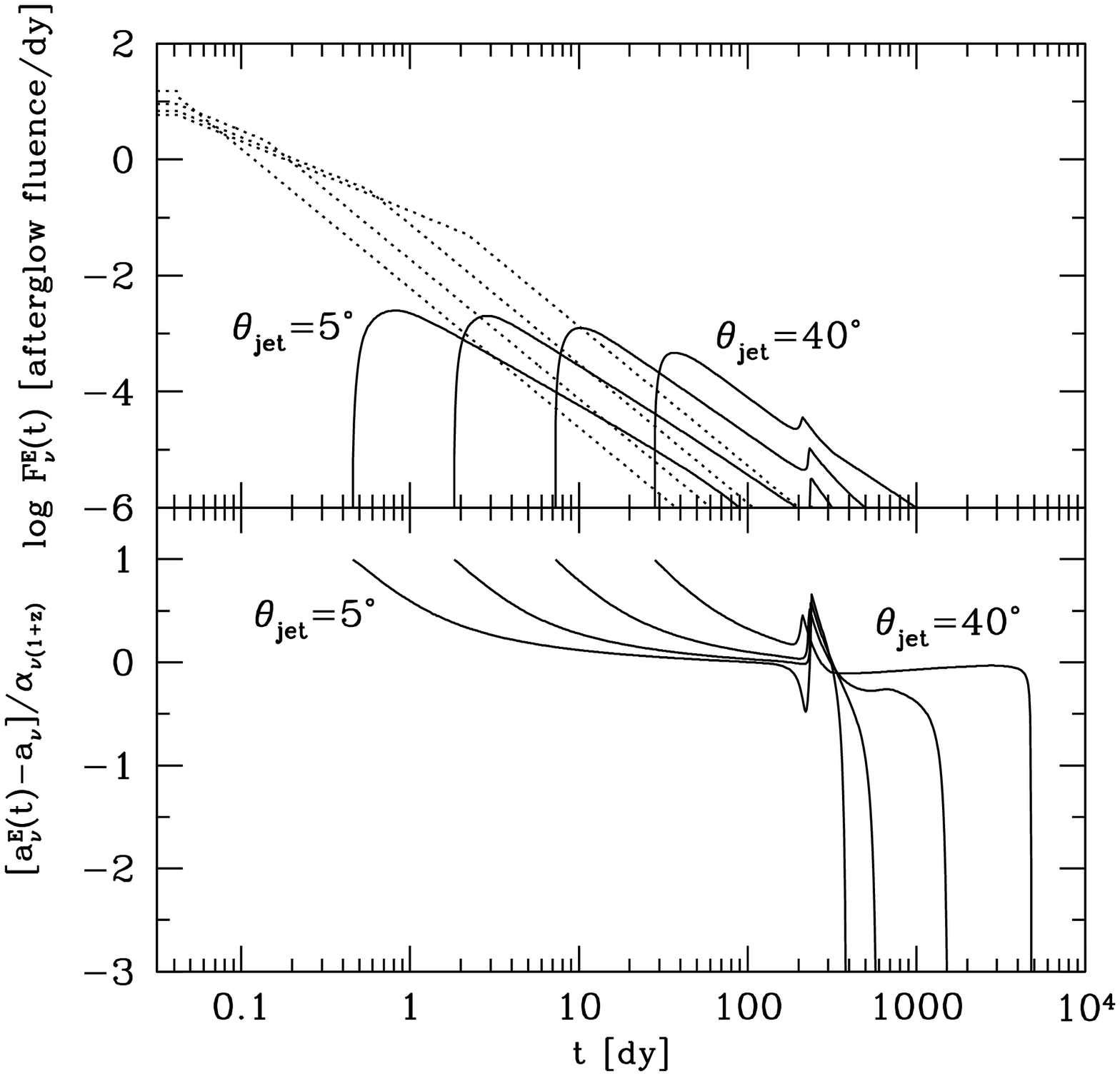]{Same as Figure 10, but with 
$[(1+z)/2][n(\rm{1\,pc})/\rm{1\,cm}^3]^{-1}(E\theta_{jet}^2/10^{52}\rm{\,erg}) = 
0.1$.  The corresponding values of $t_{jet}$ are 0.035, 0.14, 0.56, and 2.2 dy, 
respectively.  The dust echoes are again brighter than the afterglows at times 
substantially later than $t_{jet}$.}

\figcaption[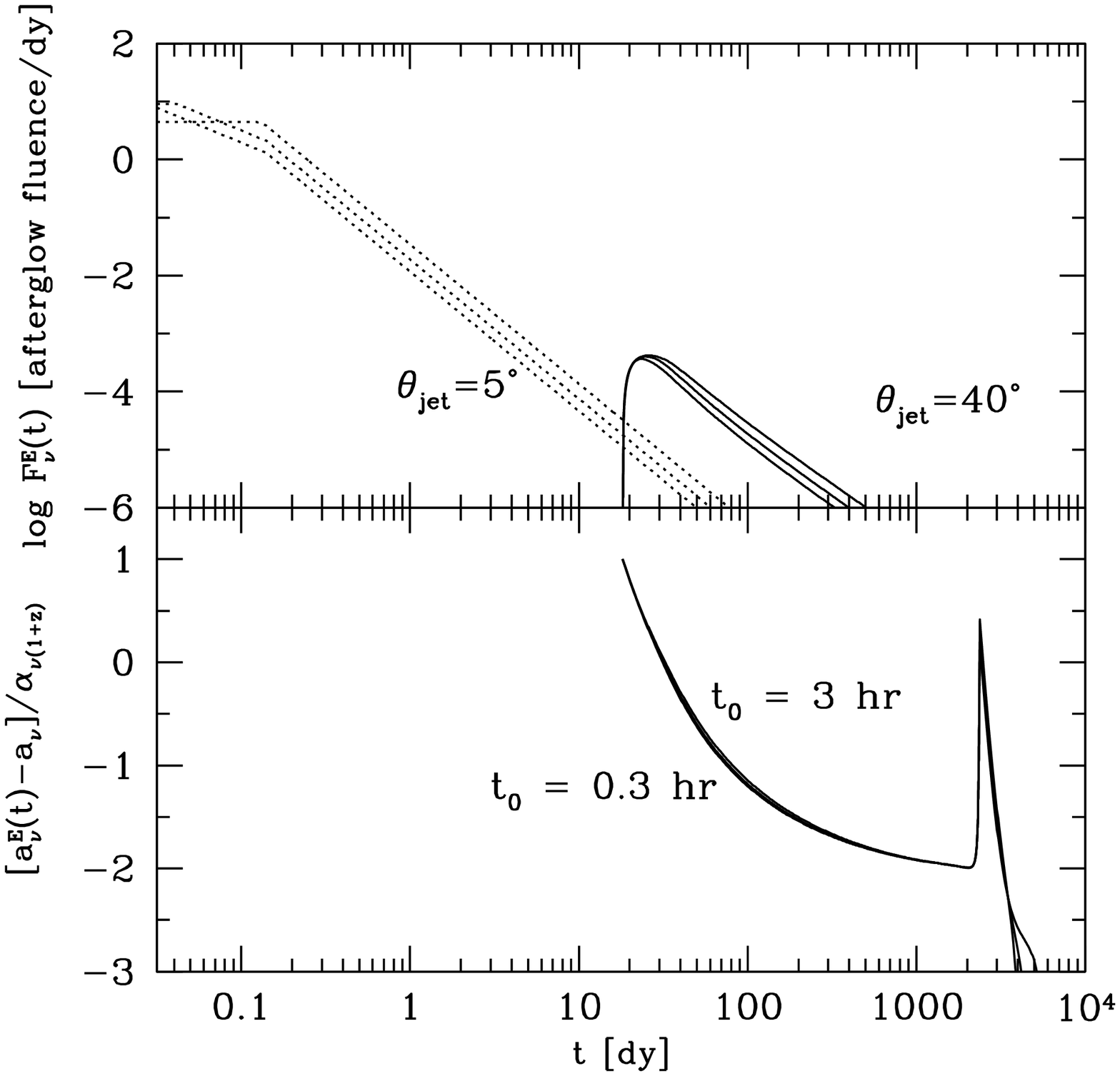]{Light curves (top panel) and color histories (bottom 
panel) of scattering dust echoes (solid lines) of afterglow light (dotted 
lines), with $f = 1$, $R(1+z) = 1$ pc, $\tau_{\nu (1+z)} = 3$, $t_0 = 0.3$, 1, 
and 3 hr, $\theta_{jet} = 10^{\circ}$, and 
$[(1+z)/2][n(\rm{1\,pc})/\rm{1\,cm}^3]^{-1}(E\theta_{jet}^2/10^{52}\rm{\,erg}) = 
0.1$.  The corresponding value of $t_{jet}$ is 0.14 dy.  Clearly, the choice of 
$t_0$ has little affect on the light curve and color history of the dust echo.}

\figcaption[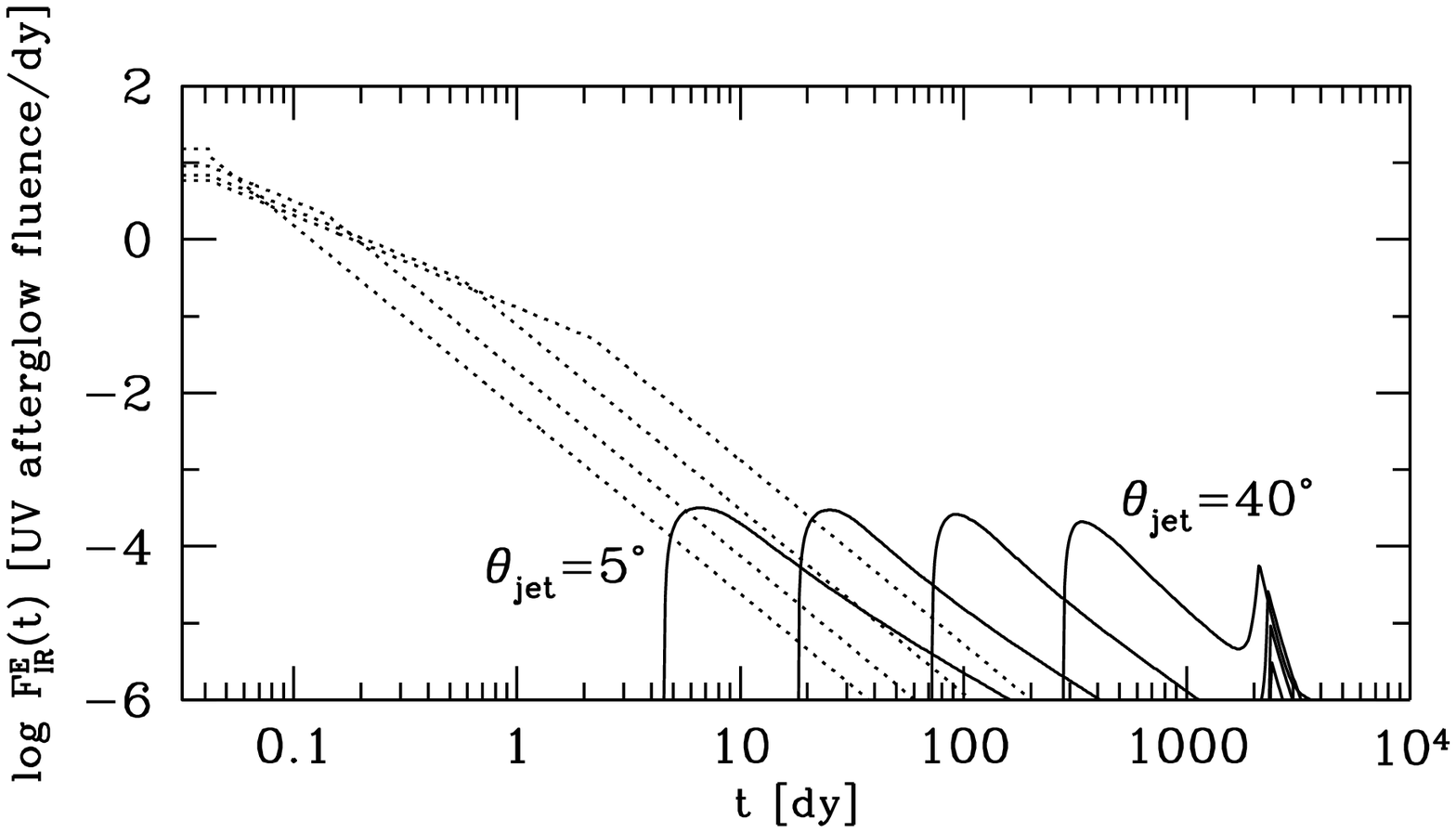]{Light curves of thermal dust echoes (solid lines) of 
afterglow light (dotted lines), with $f = 1$, $R(1+z) = 1$, $\tau_{\nu (1+z)} = 
3$, $t_0 = 1$ hr, $\theta_{jet} = 5^{\circ}$, $10^{\circ}$, $20^{\circ}$, and 
$40^{\circ}$, and 
$[(1+z)/2][n(\rm{1\,pc})/\rm{1\,cm}^3]^{-1}(E\theta_{jet}^2/10^{52}\rm{\,erg}) = 
0.1$.  The corresponding values of $t_{jet}$ are 0.035, 0.14, 0.56, and 2.2 dy, 
respectively.  Thermal dust echoes are relatively brighter at late times than 
scattering dust echoes, because large-angle changes in the direction of the 
light are just as probable as small-angle changes, but this is not the case with 
scattering.  Large-angle changes are especially important at late times.}

\setcounter{figure}{0}

\begin{figure}[tb]
\plotone{echo1.eps}
\end{figure}

\begin{figure}[tb]
\plotone{echo2.eps}
\end{figure}

\begin{figure}[tb]
\plotone{echo3.eps}
\end{figure}

\begin{figure}[tb]
\plotone{echo4.eps}
\end{figure}

\begin{figure}[tb]
\plotone{echo5.eps}
\end{figure}

\begin{figure}[tb]
\plotone{echo6.eps}
\end{figure}

\begin{figure}[tb]
\plotone{echo7.eps}
\end{figure}

\begin{figure}[tb]
\plotone{echo8.eps}
\end{figure}

\begin{figure}[tb]
\plotone{echo9.eps}
\end{figure}

\begin{figure}[tb]
\plotone{echo10.eps}
\end{figure}

\begin{figure}[tb]
\plotone{echo11.eps}
\end{figure}

\begin{figure}[tb]
\plotone{echo12.eps}
\end{figure}

\begin{figure}[tb]
\plotone{echo13.eps}
\end{figure}

\begin{figure}[tb]
\plotone{echo14.eps}
\end{figure}

\end{document}